\documentclass[a4paper]{article}
\usepackage{epsfig}
\usepackage{psfig}
\usepackage{latexsym}
\usepackage{amsmath}
\usepackage{graphics}
\usepackage{graphpap}
\begin{document}
\title{Dynamical diffraction in sinusoidal potentials: uniform 
approximations for Mathieu functions}
\author{D H J O'Dell \\  \vspace{0.5ex} Chemical Physics Department,
 Weizmann Institute of Science, \\ 
\vspace{0.5ex} Rehovot 76100, Israel \\ \vspace{0.5ex}
E-mail: \textsf{duncan.o'dell@weizmann.ac.il}}
\maketitle
\begin{abstract}
\textbf{Abstract.}  Eigenvalues and eigenfunctions of  Mathieu's equation 
 are found in 
the short wavelength limit using a uniform approximation (method of comparison with a 
`known' equation having the same classical turning point structure) applied in
Fourier space. 
The uniform approximation used here relies upon the fact that by 
passing into Fourier space the Mathieu equation can be mapped onto the simpler problem
 of a double well potential.
The resulting eigenfunctions (Bloch waves), which are uniformly valid for all 
angles, are then used to describe the semiclassical scattering of waves by potentials 
varying sinusoidally in one direction.
In such situations, for instance in the diffraction of atoms by gratings made of light, it 
is common to make the Raman-Nath approximation which ignores the motion of
the atoms inside the grating. When using the eigenfunctions no such approximation is 
made so that the dynamical diffraction regime (long interaction time) can be 
explored. 
\end{abstract}

\begin{figure}[htbp]
\begin{center}
\centerline{\epsfig{figure=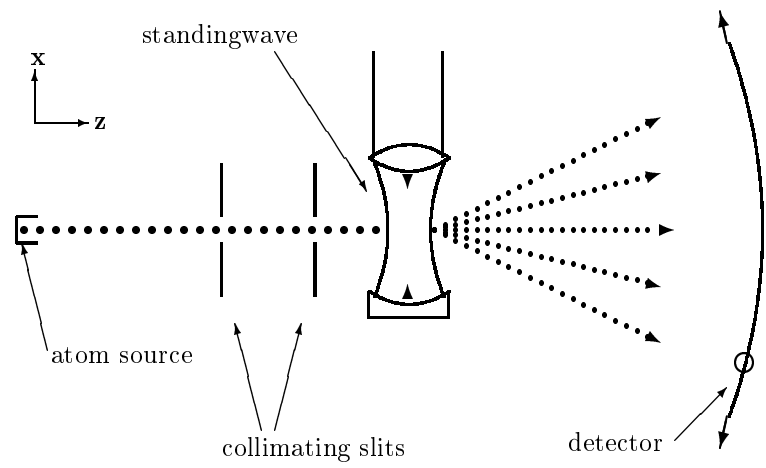,width=10cm,angle=0}}
\end{center}
\caption{A typical experimental set-up used in the investigation of atomic
diffraction.}
\label{fig:expt}
\end{figure}

\textbf{1. Introduction}
Consider a diffraction experiment in two dimensions, as depicted in Figure \ref{fig:expt}. 
A 
plane wave $\exp (ikz)$ propagates freely in the $\hat{z}$ (longitudinal) direction and is
 incident normally upon a medium with a refractive index which varies weakly in the 
$\hat{x}$ (transverse) direction
\begin{equation}
n(x)=n_{0}+n_{1} \cos 2Kx  \quad \quad (n_{1} \ll n_{0}).
\end{equation}
As the refractive index depends only on $x$, the wavefunction inside the medium separates, 
$\Psi(z,x)=\phi(z)\psi(x)$, with $\phi(z)$ being trivially given by
\begin{equation}
\phi(z)= \exp \left(\mathrm{i} z \sqrt{k^2 n_{0}^2- \kappa}  \right)
\end{equation}
with $\kappa$ a separation constant proportional to the transverse energy
of the wave inside the medium.
The transverse behaviour is governed by
Mathieu's equation (see Abramowitz and Stegun 1964)
\begin{equation}
\frac{\partial^{2} \psi (x)}{\partial x^2}
 +\left(\kappa+2k^{2}n_{0}n_{1} \cos 2Kx \right) \psi(x) = 0.
\label{eq:mathieu}
\end{equation}
More generally, Mathieu's equation is obtained whenever the 3-dimensional Helmholtz wave 
equation is separated in elliptical coordinates, useful for, say,
scattering from elliptical boundaries.
This paper is concerned with the short wavelength (semiclassical) regime
for which the parameter $2k^{2}n_{0}n_{1}/K^{2}$ is very large. One motivation is that the 
resulting asymptotics are known to describe the emergence of interesting classical features such
as caustics (singularities of the geometric ray theory) which come to dominate the wave
 field as the wavelength is reduced to zero. The caustic structure 
becomes ever more intricate as the (longitudinal) thickness of the medium is
increased (Berry and O'Dell  1999).

Diffraction by a sinusoidal grating has been studied 
in the context of the diffraction of light by ultrasound since at least
1921 (Brillouin  1921) (see Berry 
(1966) for a review to 1966). More recent interest has arisen through
the realisation of the diffraction of beams of atoms by beams of light
(Adams \textit{et al}  1994). Then 
$n(x)= \sqrt{1-V(x)/E}$, where $E$ is the energy of the atoms and $V(x)$
is the potential energy
due to their interaction with a standing wave of light 
(Cohen-Tannoudji \textit{et al} 1992, 
Kazantsev \textit{et al} 1991)
\begin{equation}
 V(x) = -\frac{\Delta}{4 \hbar} \frac{d^{2} \; \mathcal{E}_{0}^{2}}{\Delta^{2}+\Gamma^{2}/4}
\cos^{2} K x = - V_{0} \cos^{2} K x \label{eq:atom-light-pot}
\end{equation}
with $\mathcal{E}_{0}$ the magnitude of the electric field of the counter propagating laser
 beams which form the standing wave, $K$ their wavenumber and $\Delta$ the frequency
 detuning from resonance. $\Gamma$ is the
spontaneous decay rate for the excited atom and $d$ the atomic dipole moment 
for the electronic transition being used. 

The resonant nature of the atom-light
interaction allows for an efficient transfer of transverse kinetic energy
to the atoms.  Combined with their large mass, this means the atoms can attain small 
transverse de~Broglie wavelengths and so are
rather good candidates to access the semiclassical scattering regime when
compared with other microscopic particles such as neutrons or electrons. 

The idealised experiment described above assumes that
 the standing wave laser field has a `top-hat' cross-section in the longitudinal direction,
 switching on at $z=0$, and remaining constant till switching off at $z=Z$ at 
which point the atoms propagate undisturbed to a detector in the farfield. 
One might achieve this by telescopically expanding the normally
gaussian laser profile and physically masking the entry and exit edges to make
them sharp (diffraction limited). In this way the entry and exit into the
laser can be sudden from the point of view of the dynamics of the centre of mass
of the atom, but still adiabatic from the point of view of the rapid
Rabi oscillations of the internal electronic states that generate the potential
given by Equation (\ref{eq:atom-light-pot}), see O'Dell (1999) 
for more details. In any case, features such as caustics will
still be qualitatively correctly described by the simple model given here even
if experimental conditions differ considerably. This is because caustics
are stable to perturbations as guaranteed by the optical catastrophe 
theory (Berry 1980).

\textbf{2. Dynamical diffraction and the Raman-Nath equations}

The time independent
Schr\"{o}dinger equation governing the passage of atoms of energy 
$E=\hbar^{2}k^{2}/2m$ through an optical standing wave of periodicity
$\pi/K$ is
\begin{equation}
\frac{\partial^{2} \Psi}{\partial x^2}+\frac{\partial^{2} \Psi}{\partial z^2}
 +\left(k^{2} + \frac{2 m V_{0}}{\hbar^{2}} \cos^{2}(K x) \right) \Psi = 0.
\label{eq:schrod}
\end{equation}
The periodic potential suggests
 an atomic wavefunction of the form
\begin{equation}
\Psi(x,z)= \mathrm{e}^{\mathrm{i}kz} \sum_{n=- \infty}^{\infty} A_{n}(z)
 \mathrm{e}^{2\mathrm{i}nKx}. \label{eq:wavefunction-ansatz}
\end{equation}
An advantage of this decomposition of the wavefunction is that upon exiting the
 interaction region at $z=Z$, the terms in Equation 
(\ref{eq:wavefunction-ansatz}) represent freely propagating diffracted
waves travelling at angles $\arcsin (2nK/k)$ to the $z$ axis with amplitudes
$A_{n}(z=Z)$. A detector in the farfield will register a diffraction
pattern made up of discrete beams with intensities $|A_{n}(Z)|^{2}$. The amplitudes
 satisfy $\sum_{n=-\infty}^{n=+\infty}|A_{n}|^{2}=1$. The rest of this paper is
 devoted to determining the $A_{n}(z)$.

If the initial kinetic energy of the atoms is thermal then $k \gg  K$ and
the propagation of the atom beam is paraxial. The paraxiality means that the evolution
of the $A_{n}$ with $z$ will be much slower than $\exp (ikz)$ so when
substituting 
(\ref{eq:wavefunction-ansatz}) into (\ref{eq:schrod}) terms
containing $d^{2} A_{n}/d z^{2}$ can be ignored. The result is an 
infinite series of coupled equations
\begin{equation}
\mathrm{i}\frac{\partial A_{n}}{\partial \zeta}-n^{2}A_{n} + 
\frac{\Lambda}{2} \left(A_{n+1}+2A_{n}+A_{n-1} \right)=0
\end{equation}
where
\begin{equation}
\zeta  \equiv  \frac{2K^{2}z}{k} \quad , \quad
\Lambda  \equiv  \frac{mV_{0}}{4\hbar^{2}K^{2}}. \label{eq:lambda-definition}
\end{equation}
The parameter $\Lambda$ is equivalent to the parameter $2k^{2}n_{0}n_{1}/K^{2}$
 appearing above in Eq. (\ref{eq:mathieu})---by letting $\hbar \rightarrow 0$,
and hence $\Lambda \rightarrow \infty$, one obtains the classical limit.
 Of course, taking the limit 
$\hbar \rightarrow 0$ is a formal device. In an actual experiment the 
short wavelength limit is approached by, say, making $V_{0}$, the interaction between
the atoms and the light, as large as possible. This increases the depth of the
wells of the sinusoidal potential which in turn means there are more quantised
transverse states.
A phase transformation $A_{n} \rightarrow A_{n} \exp(\mathrm{i} \Lambda \zeta)$ 
slightly simplifies the equations to
\begin{equation}
\mathrm{i}\frac{\partial A_{n}}{\partial \zeta}-n^{2} A_{n} + 
\frac{\Lambda}{2} \left(A_{n+1}+A_{n-1} \right)=0. \label{eq:raman-nath}
\end{equation}
These are (apart from a straight forward
change of variables) the differential difference equations introduced by Raman and Nath  
(1935,1936) to describe the diffraction
of light by ultrasound. The Raman-Nath (RN) equations are a description
of dynamical diffraction, yielding the evolution of the amplitudes
of the various diffracted beams as the atom wave passes through the light
grating. In their original paper, Raman and Nath (1935)
observed that by ignoring the diagonal term, $n^{2} A_{n}$, one obtains
simple solutions for $A_{n}$ in terms of Bessel functions. This is
equivalent to neglecting the transverse kinetic energy of the atoms and the sinusoidal
potential then acts only as a pure phase grating, see Berry  (1966).
 This is a very successful approximation for short interaction times  
(Sanders 1936, Gould \emph{et al} 1986, Rasel \emph{et al} 1995)
 but
\emph{dynamical} diffraction requires that full account be taken of the transverse motion.

A general property of paraxial systems is that the axial coordinate, in this case 
the rescaled longitudinal distance, $\zeta$, plays the r\^{o}le of time. 
The numerical integration of the RN equations is relatively simple since they
 are first order differential equations in $\zeta$ with only a single
boundary condition: $A_{n}(\zeta=0)=\delta_{n0}$. However, when investigating
the behaviour at long interaction times it becomes more economic to analyse the
problem in terms of the eigenfunctions of the scattering potential,
which propagate unchanged through the medium. This approach will be adopted by
 seeking eigenfunctions of the RN equations (\ref{eq:raman-nath}) of the
 form $A_{n} \equiv B_{n} \exp (-\mathrm{i} E \zeta)$. Corresponding to each
 eigenvalue $E^{j}$ is an eigenfunction consisting of a `vector' of amplitudes 
$(B_{-\infty}^{j},\ldots,B_{-2}^{j},B_{-1}^{j},
B_{0}^{j},B_{1}^{j},B_{2}^{j},\ldots,B_{\infty}^{j})$,
whose elements satisfy
\begin{equation}
E^{j} B_{n}^{j}= n^{2} B_{n}^{j} - \frac{\Lambda}{2}
 \left(B_{n+1}^{j}+B_{n-1}^{j} \right). \label{eq:R-N-stationary}
\end{equation}
This equation defines the tridiagonal RN matrix hamiltonian. For a weak
 potential (i.e.\ small $\Lambda$: the quantum, non-classical limit) the RN
matrix can be approximated by a $3 \times 3$, or for the special case
 of oblique incidence near a Bragg angle, a $2 \times 2$ matrix,
 and analytical solutions for the Bloch waves (eigenfunctions)
 are easy to find (Berry and O'Dell  1998).
 Here, however, we are interested in the opposite limit.

For the purposes of numerical diagonalisation, a guide to the minimum
 diffraction order, $\pm N$, at which the RN matrix can be safely
 truncated for large $\Lambda$ is given by 
\begin{equation}
N \; = \; \sqrt{2 \Lambda} \; \; \propto \hbar^{-1}. \label{eq:trunc}
\end{equation}
This includes only those beams contained within the maximum scattering
 angle that can be achieved \emph{classically} (Berry 1966).
 Criterion (\ref{eq:trunc}) becomes exact as
 $\Lambda \rightarrow \infty$ but at the cost of requiring an 
infinite number of beams.

The `physical' derivation of the stationary RN equation
(\ref{eq:R-N-stationary}) given here is nothing more than
a Fourier analysis of Mathieu's equation (\ref{eq:mathieu}). Equation
(\ref{eq:R-N-stationary}) is the recursion relation 
satisfied by the Fourier coefficients of periodic solutions to
 Mathieu's equation: the even and odd \emph{Mathieu functions}
 (Abramowitz and Stegun  1964). The eigenvalues $E^{j}$
 are known as the \emph{characteristic values}.

\textbf{3. WKB solution of the Raman-Nath equations}

In the semiclassical limit the dimensions of RN matrix defined by Equation 
(\ref{eq:R-N-stationary}) are infinite and numerical diagonalisation
becomes impossible. Dingle and Morgan 
 (1967a,1967b) (see appendix of Berry 1966) and
 independently Yakovlev (1997), have given a
 WKB-type solution for the \emph{continuised} RN equations accurate in
 the large $\Lambda$ limit. The idea
is to capture the very fast oscillation of the Bloch wave 
(see Figures \ref{fig:uniform0}--\ref{fig:uniform200})
 with the exponential of a slowly varying function. To this end one replaces
 the discrete variable $n$ with the continuous one $y$
\begin{equation}
y  \equiv  \frac{n}{\sqrt{\Lambda}} \label{eq:y-definition}
\end{equation}
so that when $n \rightarrow n+1$, then $y \rightarrow y +
 (\sqrt{\Lambda})^{-1}$
and the discrete amplitudes become continuous functions of $y$,
 $B_{n} \rightarrow B(y)$.  Defining the rescaled eigenvalue
\begin{equation}
\beta  \equiv  \frac{E}{\Lambda} \label{eq:beta-definition}
\end{equation}
the stationary RN equation (\ref{eq:R-N-stationary}) becomes
\begin{equation}
(\beta-y^{2})B(y)+\frac{1}{2} \left[B \left(y+(\sqrt{\Lambda})^{-1} \right)+B
 \left(y-(\sqrt{\Lambda})^{-1} \right) \right]=0. \label{eq:rn-stationary-scaled}
\end{equation}
Following Berry (1966), let
\begin{equation}
B(y)=\mathrm{e}^{\mathrm{i}S(y)}
\end{equation}
where the suggestively named $S(y)$ is analogous to an action. Taylor expanding
$S(y+ (\sqrt{\Lambda})^{-1})$ and $S(y-(\sqrt{\Lambda})^{-1})$ gives a differential
 equation of infinite order
\begin{equation}
\cos \left( S^{i}+ \frac{S^{iii}}{6}+ \cdots \right)= (y^{2}-
\beta)\mathrm{e}^{-\mathrm{i}(S^{ii} \! / \! 2 \:+S^{iv} \! / \! 24 
\: + \cdots)}
\end{equation}
where $S^{m}=(\sqrt{\Lambda})^{-m} \; \partial^{m} S/\partial y^{m}$ 
are assumed to be small quantities of order $(\sqrt{\Lambda})^{-m}$.
Solving for the first derivative one has
\begin{equation}
\frac{1}{\sqrt{\Lambda}}\frac{\partial S}{\partial y} = \arccos 
\left[(y^{2}-\beta)\mathrm{e}^{-\mathrm{i}(S^{ii} \! / \! 2 \:+S^{iv} 
\! / \! 24 \: + \cdots)} \right]- \frac{S^{iii}}{6}- \cdots. 
\label{eq:rn-wkb-exact}
\end{equation}
Expanding the right hand side (rhs) gives
\begin{equation}
\begin{split}
\frac{1}{\sqrt{\Lambda}}\frac{\partial S}{\partial y}= \arccos  \; 
[y^{2}-\beta]  & + \mathrm{i}\frac{y^{2}-\beta}{\sqrt{1-(y^{2}-
\beta)^{2}}} \frac{S^{ii}}{2}  - \frac{S^{iii}}{6} + 
\mathrm{i}\frac{y^{2}-\beta}{\sqrt{1-(y^{2}-
\beta)^{2}}}\frac{S^{iv}}{24}\\ & + \frac{y^{2}-\beta}{\left(1-
(y^{2}-\beta)^{2} \right)^{3/2}}\frac{(S^{ii})^{2}}{8}+ \cdots
\end{split}
\end{equation}
which can be solved for $S^{i}$ by iteration, yielding to second order 
\begin{equation}
\frac{\partial S}{\partial y} \approx \sqrt{\Lambda}\;  \arccos \; [y^{2}-
\beta] -\mathrm{i}\frac{(y^{2}-\beta)y}{1-(y^{2}-\beta)^{2}}.
\end{equation}
The second term on the rhs can be integrated immediately so that
the equation for $S$ can be written
\begin{equation}
\begin{split}
S & \approx \sqrt{\Lambda}\; \int  \arccos \; [y^{2}-\beta] \; dy -
\frac{\mathrm{i}}{4} \ln \left(1-(y^{2}-\beta)^{2} \right) \\
& \equiv \sqrt{\Lambda}\; S_{0}(y,\beta) -
\frac{\mathrm{i}}{4} \ln \left(1-(y^{2}-\beta)^{2} \right).
\end{split}
\end{equation}
Thus the continuised eigenfunctions take the form (Berry 1966)
\begin{equation}
B(y)=\mathrm{e}^{\mathrm{i} 
S} \approx \frac{\mathrm{e}^{\mathrm{i}\sqrt{\Lambda}\; \int  \arccos \; 
[y^{2}-\beta] \; dy}}{\left(1-(y^{2}-\beta)^{2} \right)^{1/4}} = 
 \frac{\mathrm{e}^{\mathrm{i}\sqrt{\Lambda}\;
 S_{0}(y,\beta)}}{\left(1-(y^{2}-\beta)^{2} \right)^{1/4}}
\label{eq:wkb}
\end{equation}
which resembles a WKB expression.
In particular the denominator causes divergences at the turning-points
\begin{equation}
y= \pm \sqrt{\beta \pm 1}.
\end{equation}
`Bound' solutions of Mathieu's equation have energies lying between the top
and bottom of the sinusoidal wells ($0 \leq E \leq V_{0}$) which translates into
the bound eigenvalues occupying the range $-1 \leq \beta \leq 1$. This
 means that except for the situation when $\beta=1$, \emph{real}
turning points for bound states are located at
\begin{equation}
y_{\pm}= \pm \sqrt{\beta + 1}.
\end{equation}
`Free' states have energies above the wells. When $E$ is sufficiently
 greater than $V_{0}$ it is easy to find WKB solutions
of Mathieu's equation directly in coordinate space (since they have no
 classical turning points and hence no divergences) rather than the
 momentum (Fourier) space used here. Similarly, it is also simple to
 find eigenfunctions in coordinate space
when $\beta \approx -1$, since states near the very bottom of the wells
 are the most localised and see an essentially harmonic potential yielding
 hermite polynomials as solutions. The most interesting situation is for
 $\beta$ lying  close to $+1$. These are the states affected most by
 tunnelling between the wells and will be examined in
sections 10--19. 

The remaining `action' integral, giving the phase of the WKB
 solution (\ref{eq:wkb}), 
can be calculated using the positive turning point $y_{+}= \sqrt{\beta+1}$ as
the lower limit (i.e.\ the zero or reference point of the phase)
\begin{equation}
\begin{split}
S_{0}(y_{+},y,\beta) & =\int_{\sqrt{\beta + 1}}^{y}  \arccos \;
 [{y'}^{2}-\beta] \; dy' \\ & = 
y \arccos \;[y^{2}- \beta]-2 \sqrt{1+\beta} \; \mathrm{E} 
\left(\frac{1}{2} \arccos \; [y^{2}-\beta] \right| \left. 
\frac{2}{1+\beta} \right) \label{eq:wkb-phase}
\end{split}
\end{equation}
where $\mathrm{E}(\phi|m)= \int_{0}^{\phi} \sqrt{1-m \sin^{2} \theta} \, d 
\theta$ is the incomplete elliptic integral of the second kind,
 see Gradshteyn  and Ryzhik  (1965).
This expression for the phase is valid for $0 \leq y \leq \sqrt{\beta+1}$.
 For perpendicular incidence the phase is symmetrical about $y=0$,
 and so only this half-range is required. For values of $y$ greater
 than $\sqrt{\beta+1}$ the phase is purely imaginary, but with care
 (\ref{eq:wkb-phase}) still gives the correct answer.

\textbf{4. Single and double wells in momentum space}

\begin{figure}[htbp]
\begin{center}
\centerline{\epsfig{figure=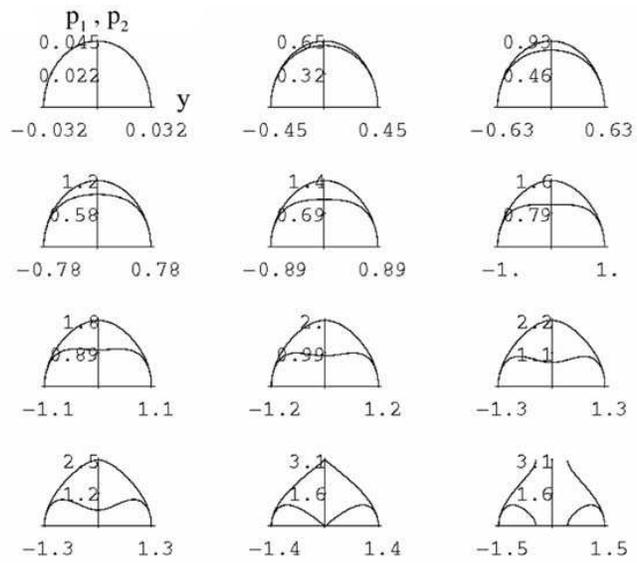,width=10cm,angle=0}}
\end{center}
\caption{A series of plots showing the two momentum 
functions, $p_{1}$ and $p_{2}$, as functions of $y$ for different 
values of $\beta$. The top left has $\beta=-0.999$, each successive 
picture has $\beta$ increasing by $0.2$ until the bottom right which
has $\beta=1.201$. It is the $p_{1}$ curve that dips down
to zero when $\beta=1$.} \label{fig:two-momenta}
\end{figure}

\begin{figure}[htbp]
\begin{center}
\centerline{\epsfig{figure=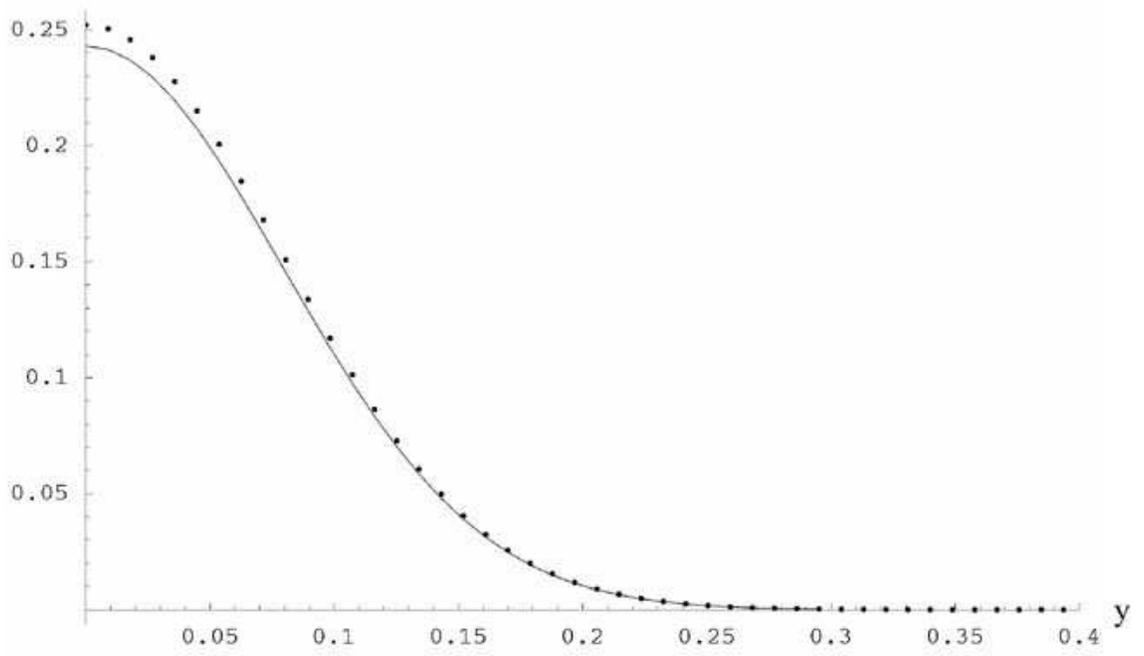,height=15cm,angle=-90}}
\end{center}
\caption{A comparison of the numerical (dots)
$0^{\mathrm{th}}$ Bloch wave,  out of 200 bound states, 
of the R-N matrix (\ref{eq:R-N-stationary}) for $\Lambda=12500$, 
with its uniform approximation (solid line). $\beta=-0.9937$.
 Since the Bloch wave is symmetrical about $y=0$, only the 
positive half is shown.} 
\label{fig:uniform0}
\end{figure}

\begin{figure}[htbp]
\begin{center}
\centerline{\epsfig{figure=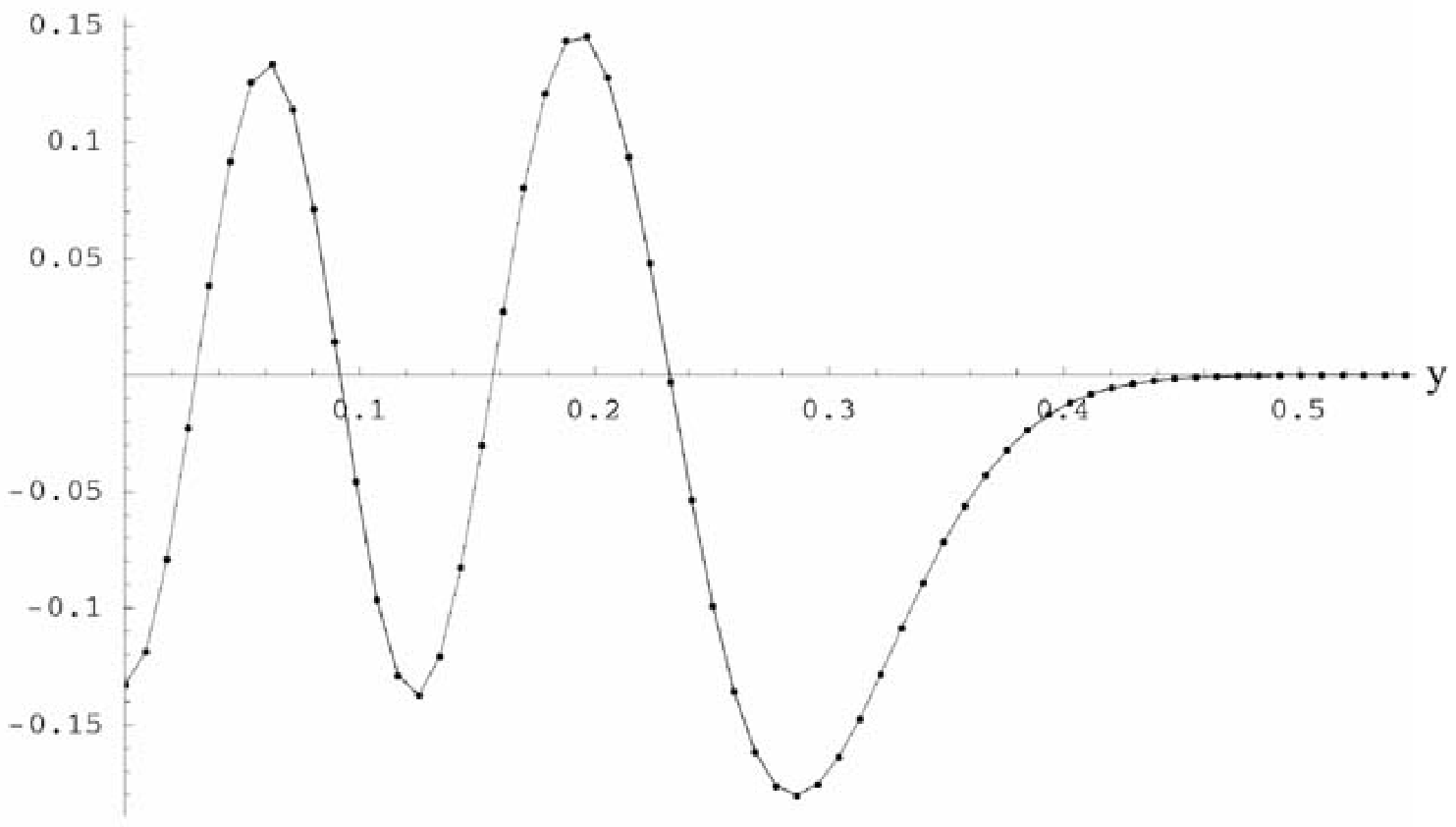,height=15cm,angle=-90}}
\end{center}
\caption{A comparison of the numerical  (dots)
$8^{\mathrm{th}}$ Bloch wave,  out of 200 bound states, 
of the R-N matrix (\ref{eq:R-N-stationary}) for $\Lambda=12500$, 
with its uniform approximation (solid line). $\beta=-0.8932$.} 
\label{fig:uniform8}
\end{figure}

\begin{figure}[htbp]
\begin{center}
\centerline{\epsfig{figure=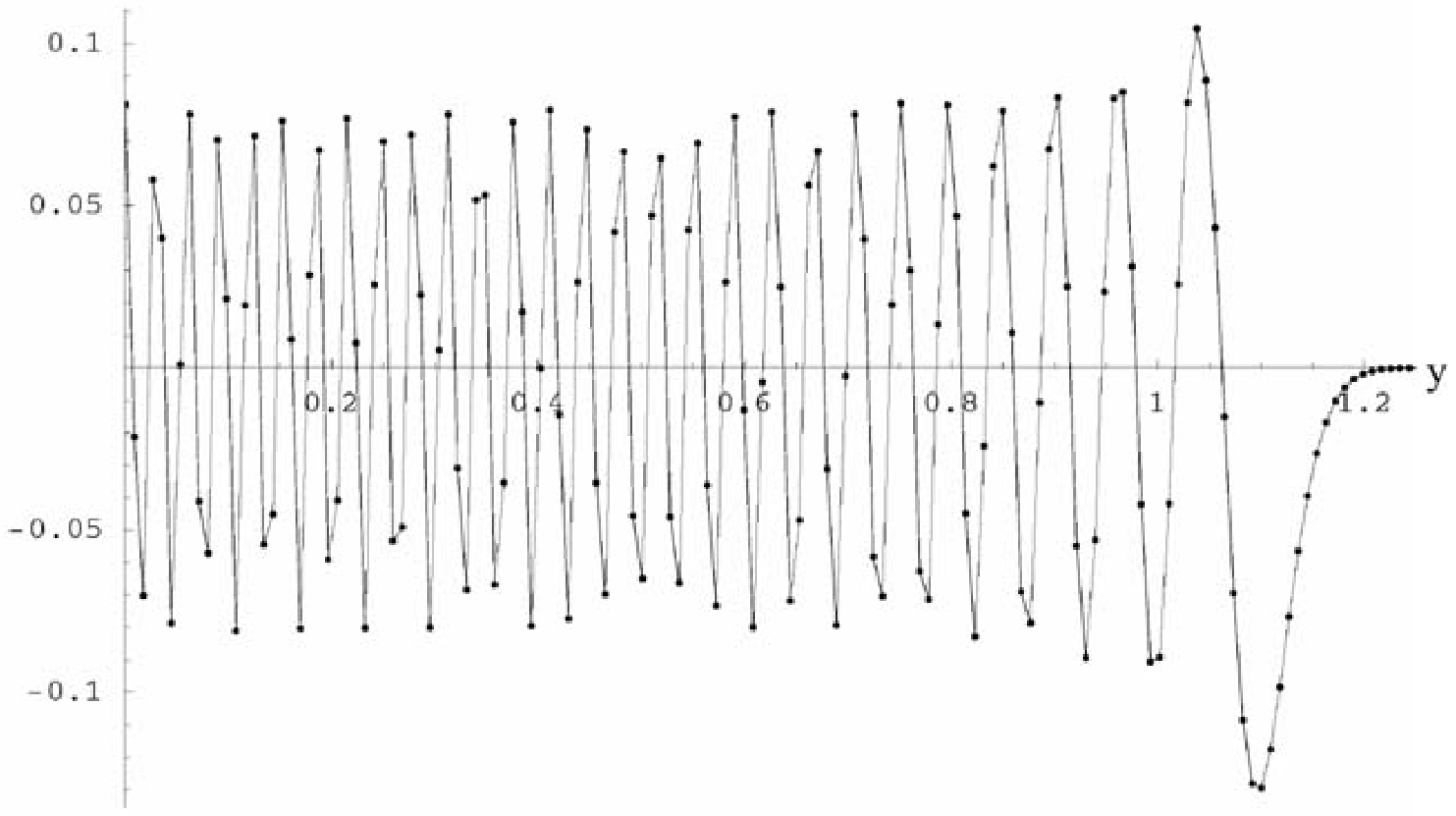,height=15cm,angle=-90}}
\end{center}
\caption{A comparison of the numerical (dots) $110^{\mathrm{th}}$
Bloch wave,  out of 200 bound states, 
of the R-N matrix (\ref{eq:R-N-stationary}) for $\Lambda=12500$, 
with its uniform approximation (solid line). $\beta=0.2616$.} 
\label{fig:uniform56}
\end{figure}

\begin{figure}[htbp]
\begin{center}
\centerline{\epsfig{figure=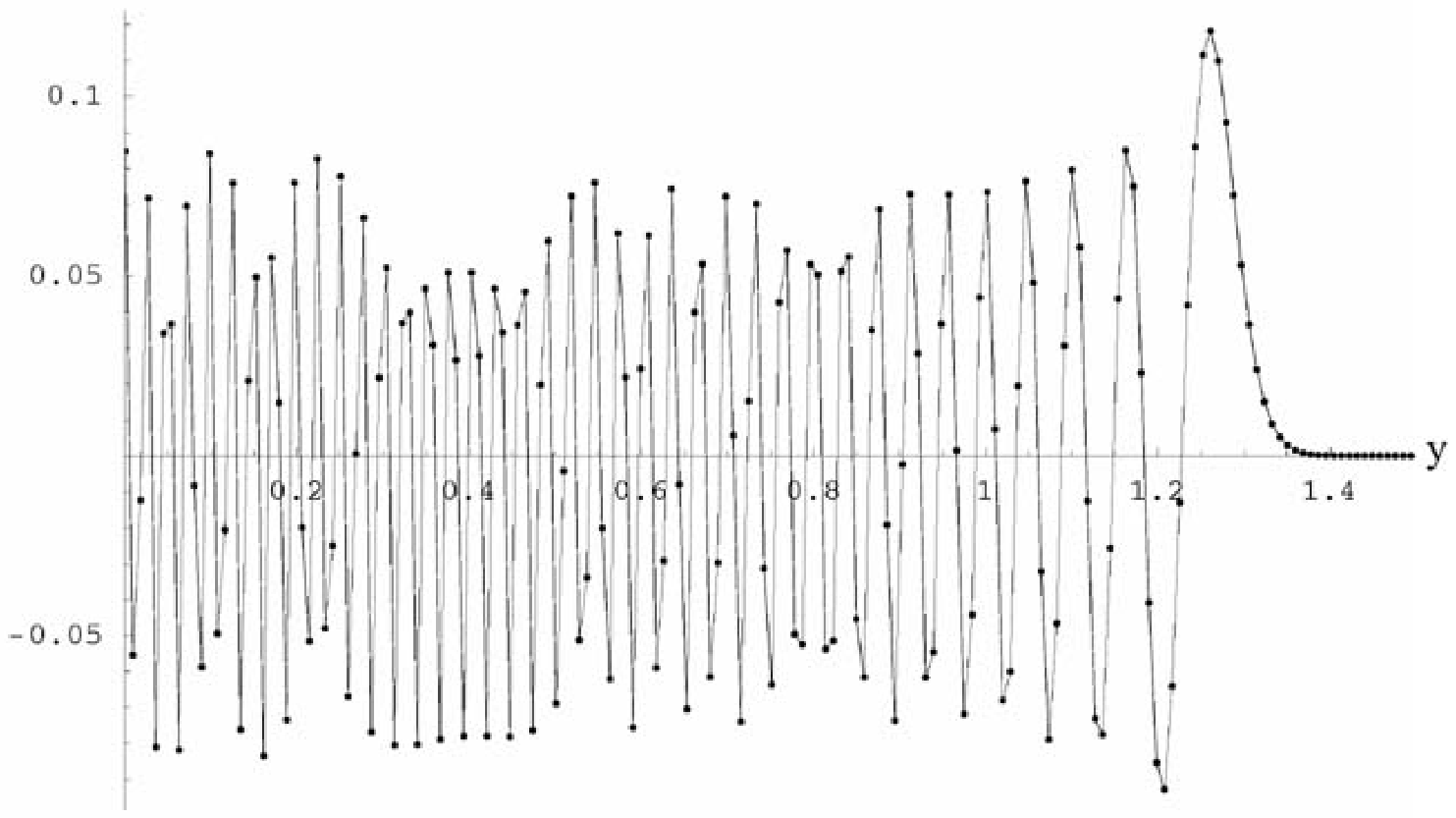,height=15cm,angle=-90}}
\end{center}
\caption{A
comparison of the numerical (dots) $152^{\mathrm{nd}}$ Bloch wave,
out of 200 bound states, 
of the R-N matrix (\ref{eq:R-N-stationary}) for $\Lambda=12500$, 
with its uniform approximation (solid line). $\beta=0.6532$.} 
\label{fig:uniform152}
\end{figure}

\begin{figure}[htbp]
\begin{center}
\centerline{\epsfig{figure=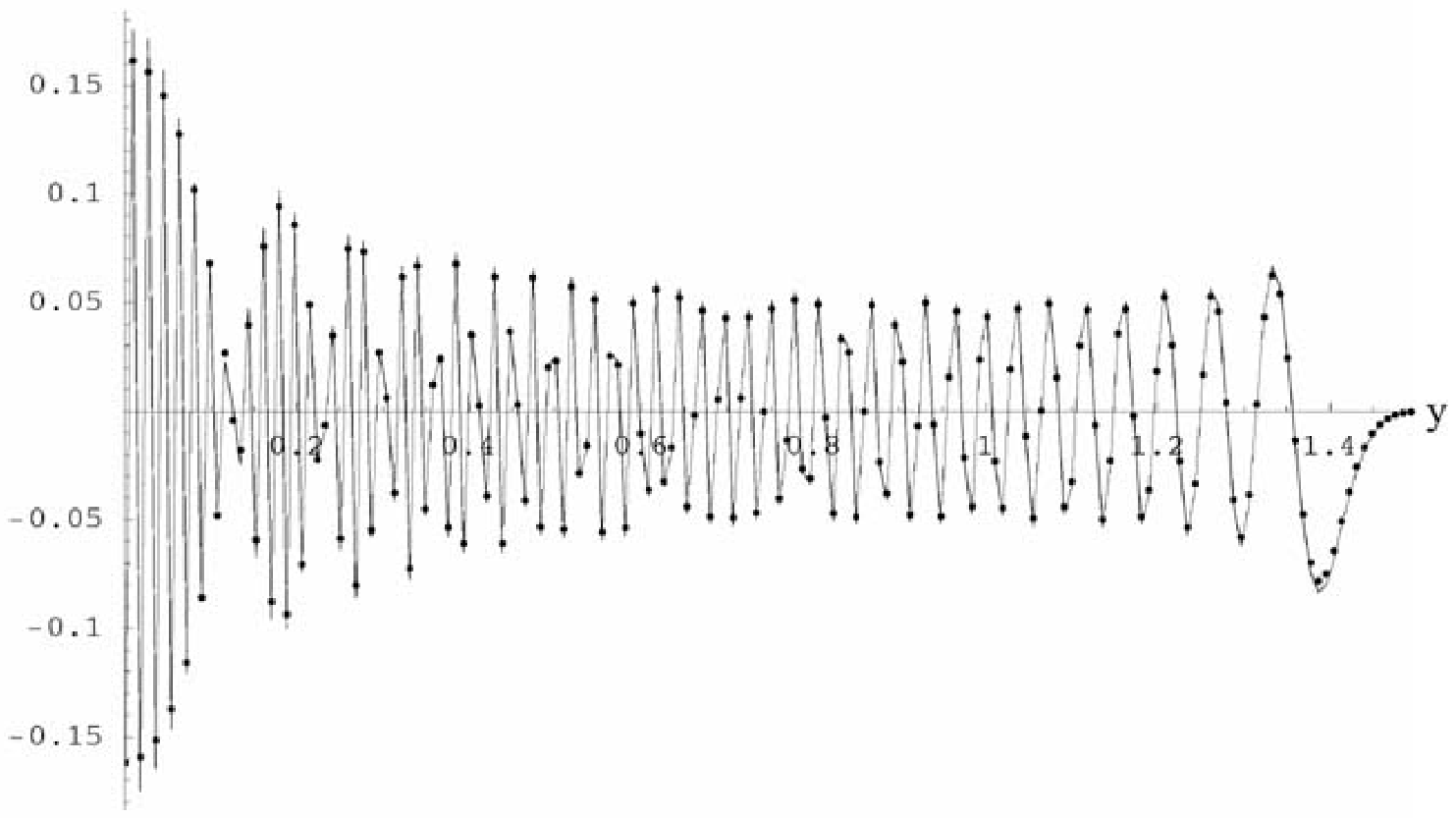,height=15cm,angle=-90}}
\end{center}
\caption{A
comparison of the numerical (dots) $200^{\mathrm{th}}$ Bloch wave,
out of 200 bound states, 
of the R-N matrix (\ref{eq:R-N-stationary}) for $\Lambda=12500$, 
with its uniform approximation (solid line). $\beta=0.9961$.} 
\label{fig:uniform200}
\end{figure}

Whilst it is useful to think of Equation (\ref{eq:wkb}) as a WKB 
type expression it does have some unusual features due to its 
unorthodox derivation from a difference equation.
Usually, for the 
Schr\"{o}dinger equation 
\begin{equation}
\frac{d^{2} \psi (q)}{d q^{2}}+ \frac{p^{2}(q)}{\hbar^{2}} \psi(q)=0
\end{equation}
where $p(q)$ is the momentum, one has the approximate WKB solution 
(Berry and Mount  1972), valid for small $\hbar$ as
 long as one is not too 
close to the turning-points ($p(q)=0$), of
\begin{equation}
\psi^{\pm}_{WKB} \equiv \frac{1}{\sqrt{p(q)}} \exp \left( \pm 
\frac{\mathrm{i}}{\hbar} \int_{0}^{q}p(q')dq' \right) \label{eq:usual-wkb}
\end{equation}
where $+/-$ refers to right/left travelling waves. Equation (\ref{eq:wkb})
 is actually for the \emph{momentum space} wavefunction, but to keep
 the analogy with the familiar coordinate space WKB solution
 (\ref{eq:usual-wkb}) simple,
Equation (\ref{eq:wkb}) will temporarily be treated as though it is a
coordinate space expression. Thus, terms such as `momentum' will refer to
functions playing the analogous r\^{o}le to $p(q)$ above. 
In particular, what is peculiar about solution (\ref{eq:wkb}) is that the
 `momentum' function appearing in the amplitude and phase are different. The two 
momenta, 
\begin{equation}
p_{1}(y,\beta) \equiv \sqrt{1-\left(y^{2}-\beta \right)^{2}} \label{eq:p1}
\end{equation}
and
\begin{equation}
p_{2}(y,\beta) \equiv \arccos \; [y^{2}-\beta] \label{eq:p2}
\end{equation}
coincide for $\beta \rightarrow -1$, but are quite different when 
$\beta \rightarrow 1$. Examining Figure \ref{fig:two-momenta}
one notes that the momentum appearing in the phase, $p_{2}$, is 
that exhibited by a particle in a simple well. The amplitude
 momentum, $p_{1}$, however, corresponds to a particle in a
 double well---although, except for values of $\beta$ greater
 than one, the particle has enough energy to move between the
 two wells. It is the classical turning-point structure that
 is of paramount importance (see Berry and Mount 
 (1972) for a review of the WKB procedure),  so the different local
 values of the two momenta give similar behaviour when $\beta <1$.
 However, as the turning-point structure of $p_{1}$ changes from
 two to four, at $\beta=1$, one can expect a qualitatively different response.

This retrospective observation, that the RN equations in some sense describe
 a wave in a single/double well, will be exploited to find uniform solutions
 in section 7. 

\textbf{5. The Bohr-Sommerfeld condition}

For states in a well, single-valuedness of the wavefunction
 dictates that only certain discrete energies are allowed.
 These eigenvalues/characteristic numbers  ensure the integral
 of the WKB phase, Equation (\ref{eq:wkb-phase}), from one turning-point
 to the other (that is, the integral across the classically accessible
 part of the potential well), correctly matches the oscillating part of
 the WKB wavefunction onto (asymptotically) exponentially decaying parts
 of the wavefunction (that tunnel into the classically forbidden sides of the well).

The action right across the well, given by setting the upper integration
 limit of Equation (\ref{eq:wkb-phase}) equal to $-\sqrt{\beta + 1}$, is also
 equal to, for perpendicular incidence, twice the value found by integrating
 only halfway, to the midpoint of the well at $y=0$,
\begin{equation}
2 S_{0}(y_{+},0,\beta)=-4 \sqrt{1+ \beta} \; \mathrm{E} \left( \frac{1}{2}
 \arccos \; [- \beta] \right| \left. \frac{2}{1+\beta} \right).
\end{equation}
The Bohr-Sommerfeld condition then states
\begin{equation}
2 \sqrt{\Lambda} \; \left| S_{0}(y_{+},0,\beta^{j}) \right|=
 \left(j+\frac{1}{2} \right) \pi \ , \qquad j=0,1,2,3 \ldots. 
\label{eq:bohr-sommerfeld}
\end{equation}
The root of this equation (which must be found numerically) for each
 value of $j$ gives the $\beta^{j}$ eigenvalues. This expression also gives
 the number of bound states that exist once a value for $\Lambda$ has been
 chosen. The term `bound' here refers to the coordinate space states that 
energetically lie below the maxima of the sinusoidal potential. In momentum
 space all the states are trapped in a well.
The most energetic bound state, labelled $j_{\mathrm{max}}$, has the value 
of $\beta$ which is closest to one. When $\Lambda$ is large the eigenvalues 
lie very close to each other, and in particular,
\begin{equation}
\lim_{\Lambda \rightarrow \infty} \beta^{j_{\mathrm{max}}}=1
\end{equation}  
and since
$\mathrm{E} \left. \left(\frac{\pi}{2}  \right|  1 \right)=1$,
then
\begin{equation}
\lim_{\Lambda \rightarrow \infty} j_{\mathrm{max}}= \frac{4 \sqrt{\Lambda}
 \sqrt{2}}{\pi} -\frac{1}{2}.
\end{equation}
Although this gives the total number of possible bound states, not all of them
 are necessarily used in the superposition which gives the diffracted wavefunction.
 The superposition coefficients, which establish the relative contribution of
 each Bloch wave to the total wavefunction, are determined from the overlap of
 the initial plane wave ($A_{n}(\zeta=0)=\delta_{n0}$) with the Bloch waves. The 
overlap integral is trivial in momentum space. 
Each coefficient is given by the value of the corresponding Bloch wave at $y=n=0$.
 Thus only the even Bloch waves are excited.  

\textbf{6. Real eigenvectors and normalisation}

Since the stationary RN equation (\ref{eq:rn-stationary-scaled}) is real it is
 always possible to find real solutions. This is achieved by constructing a 
superposition of the two independent solutions: the right and left travelling
 waves of Equation (\ref{eq:usual-wkb}), which correctly matches the exponential
 decay into the classically forbidden regions (Berry and Mount 
 1972). In the classically allowed region the single well has the solution 
\begin{equation}
B_{\mathrm{WKB}}(y,\beta)= \frac{\mathcal{N}(\beta)}{\left(1-
 \left(y^{2}-\beta \right)^{2} \right)^{1/4}} \cos \left( \sqrt{\Lambda} \;
 S_{0}(y_{+},y,\beta) + \frac{\pi}{4} \right) \label{eq:real-wkb}
\end{equation}
where $S_{0}$ is given by Equation (\ref{eq:wkb-phase}),
 and $\mathcal{N}(\beta)$ is a normalisation factor.

Normalisation of the discrete amplitudes $B_{n}$ requires that 
$\sum_{n=-\infty}^{\infty} \left| B_{n} \right|^{2}=1$.
When moving from a summation to the integration in the continuous variable $y$, care
 must be taken to include a factor of $\sqrt{\Lambda}$ which comes from the
 definition (\ref{eq:y-definition}) of $y$, so that
\begin{equation}
\sum_{n=-\infty}^{\infty} \longrightarrow \int_{n=-\infty}^{\infty} dn \longrightarrow  \sqrt{\Lambda}\int_{y=-\infty}^{\infty} dy.
\end{equation}
And so normalising the eigenvectors requires the evaluation of
\begin{equation}
\int_{-\infty}^{\infty} \left| B_{\mathrm{WKB}}^{j} (y) \right|^{2} \;dy= 1. 
\end{equation}
When $\Lambda$ becomes large the exponential decay of the wavefunction into the sides
 of the well becomes very rapid and one can ignore these contributions, so the integral
 is taken to be just that between the two turning-points of the classical motion.
 Further, the oscillation of the wavefunction is assumed to be very rapid (actually
 the normalisation factor derived in this way works well even for the ground state
 Bloch wave which has the shape of a Gaussian, i.e.\ is non-oscillatory) in
 comparison to the slow variation of the square of the amplitude. And so, without
 much loss of accuracy, the $\cos^{2}$ term  can be replaced by its average value 
of one half.  Although the amplitude diverges at the turning-points this divergence
 is still integrable. Thus, one takes
\begin{equation}
\frac{\sqrt{\Lambda}}{2} \int_{y_{-}}^{y_{+}}\frac{dy}{\sqrt{1-(y^{2}-\beta)^{2}}} 
\approx \frac{1}{\mathcal{N}^{2}}
\end{equation}
which gives
\begin{equation}
\mathcal{N}^{2} \approx \frac{\sqrt{2}}{\sqrt{\Lambda}\ 
 \mathrm{K}\left(\frac{1+\beta}{2} \right)} \label{eq:normalisation-constant}
\end{equation}
where $\mathrm{K}(m)= \int_{0}^{\pi/2} 
 (1-m \sin^{2} \theta)^{-1/2} \; d\theta$ is the complete elliptic
 integral of the first kind (Gradshteyn and Ryzhik  1965).

\textbf{7. A Uniform approximation for the Raman-Nath equation}

If the total
diffracted wavefunction was calculated as a sum over the WKB Bloch waves then
the diffraction pattern would contain spurious divergences at the turning points 
of each eigenfunction. This problem can be overcome using
 \emph{uniform approximations} which give smooth and uniformly accurate
 eigenfunctions. 
Uniform approximations are based upon the idea that one can express the solutions
 to an unstudied differential equation in terms of those of a 
well known, studied, differential equation provided the two 
share a similar transition point (classical turning point) structure. A review of
 the uniform method can be found in Berry and Mount (1972). Since the
method is central to the following calculations, it is reviewed in Appendix 1.
Somewhat non-standard, and to the best of the author's knowledge, novel
uniform approximations for  
 the momentum space Mathieu functions will be described in this and the following
 sections. 

 If one were to treat Mathieu's equation
 directly in coordinate space
then the infinite number of turning points due to the periodic well
 structure of the potential make a semiclassical analysis more complicated,
 see Berry (1971). Approximations
local to one or two turning points
(e.g.\ the expansions due to Sips, see Abramowitz and Stegun 1964)
must be carefully joined together to give the complete solution.
By using an imaginary coordinate
the Mathieu equation is converted to the modified Mathieu equation,
Abramowitz and Stegun (1964),
with a hyperbolic cosine potential, and so a uniform approximation
for a single well could be applied  (see for example 
Ancey, Folacci and Gabrielli (2000), which draws on the seminal contribution
by Olver (1954), for uniform approximations to the modified Mathieu equation
in coordinate space). 
However, for the type of diffraction
problem treated here a uniform approximation made directly in Fourier space
is more useful for two reasons. Firstly, if it is the farfield diffraction
pattern that is required, then this is the Fourier space description 
and no further transformation is required.
Secondly, for the incident plane wave initial condition considered here, 
the calculation of the superposition coefficient for each 
eigenfunction contributing to the total
wavefunction is, as described earlier, trivial.

When constructing a uniform approximation to the continuised stationary RN equation
(\ref{eq:rn-stationary-scaled}) it is not obvious what `momentum' (the pretense
of being in coordinate space rather than momentum space will continue to be
maintained)
function (i.e.\ $p(q)=\sqrt{E-V(q)}$ in Equation (\ref{eq:usual-wkb})) to use. 
The WKB-type solutions (\ref{eq:wkb}) continue to play an important r\^{o}le
 since they suggest using the
`momentum' functions $p_{1}$ and $p_{2}$ of Equations (\ref{eq:p1}) and 
(\ref{eq:p2}). It turns out that, as in the WKB case, matches to the 
actual momentum space Mathieu eigenfunctions are found when $p_{1}$ is used
for the `amplitude' part of the uniform approximation and $p_{2}$ for the `phase' part.
The mapping function should be based upon $p_{2}$, which has the structure
of a simple well. The simplest comparison equation for a well is
\begin{equation}
\frac{d^{2} \phi}{d \sigma^{2}}+\left(t-\sigma^{2}\right) \phi = 0. \label{eq:well-eqn}
\end{equation}
The parameter $t$ depends on the energy. The `equivalent points' needed for
 Equation (\ref{eq:int[gamma]dsigma=int[chi]dq}) are chosen to be the 
turning-points. This of course immediately satisfies the requirement that the
 zeros of  $\Gamma$ and $\chi$ correspond (See Appendix). Using 
(\ref{eq:int[gamma]dsigma=int[chi]dq}), the integral across the well 
gives $t$, for then
\begin{equation}
\int_{y_{-}}^{y_{+}} \sqrt{\Lambda} p_{2}(y,\beta) \; dy
 \equiv - 2 \sqrt{\Lambda} S_{0}(y_{+},0,\beta)= 
\int_{-\sqrt{t}}^{+\sqrt{t}} \sqrt{t- \sigma^{2}} \; d \sigma =
 \frac{t \pi}{2} \label{eq:equation-for-t}
\end{equation}
where $S_{0}$ is given by Equation (\ref{eq:wkb-phase}). Once $t$, which is a
 function of $\beta$, is known, the mapping function $\sigma(y)$ can be found from
\begin{equation}
\sqrt{\Lambda} S_{0}(y_{+},y,\beta)= \int_{+
 \sqrt{t}}^{\sigma(y)}\sqrt{t- \sigma^{2}} \; d \sigma=\frac{t}{2}
 \left(\arcsin \left[ \frac{\sigma}{\sqrt{t}} \right] +
 \frac{\sigma}{\sqrt{t}} \sqrt{1-\frac{\sigma^{2}}{t}} \;
 -\frac{\pi}{2} \right). \label{eq:mapping-function}
\end{equation}
Clearly this step must be executed by numerical root finding for each value 
of $y$ which is required, i.e.\ those spaced at $1/\sqrt{\Lambda}$ intervals
 which are the angular positions of the diffracted beams.

There are two standard forms of the parabolic cylinder equation
 (Abramowitz and Stegun 1964)
\begin{equation}
\frac{d^{2} \Theta}{d g^{2}} \mp \left(\frac{g^{2}}{4} \pm a \right)
 \Theta =  0. \label{eq:parabolic-cylinder-equations}
\end{equation}
The well Equation (\ref{eq:well-eqn}) corresponds to taking the upper signs.
  The simplest independent solutions to the parabolic cylinder equations are
 an even and an odd power series. However, combinations of these two power
 series lead to another two independent solutions, the Whittaker functions,
 which for large $g$ decay or grow exponentially. The Whittaker functions
 have the correct properties to match the exponential tunnelling of the
 physical solution into the sides of the well. The Whittaker solutions to the
 well equation (\ref{eq:well-eqn}) are $D_{(t-1)/2}(\sigma \sqrt{2})$ and 
$D_{(t-1)/2}(-\sigma \sqrt{2})$. The standard theory for the potential well
 uniform approximation would then predict the form of the Bloch wavefunction,
 correct for \emph{all} $y$, to be
\begin{equation}
B(y)=\psi_{\mathrm{uniform}}=\frac{1}{2} \mathcal{N} \; 2^{1/4} \left(
 \frac{2 \mathrm{e}}{t(\beta)} \right)^{t(\beta)/4}
 \left( \frac{d \sigma(y)}{dy} \right)^{-1/2} D_{(t(\beta)-1)/2}
 \left(- \sigma(y) \sqrt{2} \right) \label{eq:parabolic-cylinder-solution}
\end{equation}
and it is noted that for perpendicular incidence the choice of
 $+\sigma$ or $-\sigma$ makes no difference (only even eigenvectors are excited).
 The prefactors ensure that this expression has the same asymptotic behaviour
 when $\sigma,y \rightarrow \infty$ as the WKB solution (\ref{eq:wkb}).

Inserting the Bohr-Sommerfeld condition (\ref{eq:bohr-sommerfeld}) into the
 equation for $t$, (\ref{eq:equation-for-t}) above, gives the value of $t$
 which corresponds to the $j^{\mathrm{th}}$ eigenvalue
\begin{equation}
t=2j+1.
\end{equation}
When the index of a Whittaker function is an integer, as here, it takes on the more familiar form
\begin{equation}
D_{j}(\sigma \sqrt{2})=2^{-j/2}H_{j}(\sigma) \mathrm{e}^{-\sigma^{2}/2}
\end{equation}
where $H_{j}$ is a Hermite polynomial. However, for the higher Bloch waves
 (e.g.\ $j$ greater than 20) it is more convenient to use,  for reasons of 
speed of computation, the Airy function approximation to the Whittaker 
functions (see Abromowitz and Stegun 1964).

\textbf{8. Modifying the amplitude}

The application of the uniform method to the WKB approximation to the RN
 equation as given above requires some adjustment. The existence of two
 momentum functions means the amplitude term of Equation
 (\ref{eq:parabolic-cylinder-solution}),
\begin{equation}
\left( \frac{d \sigma(y)}{dy} \right)^{-1/2}=
\left( \frac{t- \sigma^{2}}{\arccos^{2} \left[y^{2}-\beta \right]} \right)^{1/4}
\end{equation}
does not match the WKB behaviour (see Figure \ref{fig:two-momenta}), since
 in that expression it is $p_{1}$ that appears in the amplitude. This
 disparity is removed by the substitution
\begin{equation}
\left( \frac{t- \sigma^{2}}{\arccos^{2} \left[y^{2}-\beta \right]} \right)^{1/4}
 \longrightarrow \left( \frac{t- \sigma^{2}}{1-\left(y^{2}-\beta \right)^{2}}
 \right)^{1/4}. \label{eq:amplitude-conjecture}
\end{equation}

The uniform approximation is formulated so that $\Gamma(\sigma)$ and $\chi(y)$
 approach zero together so that the divergence inherent in the WKB solution is tamed.
For the adjusted amplitude to give sensible answers this swapping of the momentum
 expressions must still lead to the correct behaviour at the turning-points.
For $\beta < 1$, both $p_{1}$ and $p_{2}$ have the same zeros, and crucially for
 the uniform approximation they go to zero in the same way, namely as the square
 root of the distance from the zero. 

L'H\^{o}pital's rule can be used to find the limiting value of the amplitude
 (\ref{eq:amplitude-conjecture}) at the turning-point. After some calculation
one finds
\begin{equation}
\lim_{y \rightarrow \sqrt{1+\beta}} \left[ \frac{t-
 \sigma^{2}}{1-\left(y^{2}-\beta \right)^{2}} \right] =
\lim_{y \rightarrow \sqrt{1+\beta}} \left[ \frac{\Gamma}{p_{1}^{2}} \right]
 = \lim_{y \rightarrow \sqrt{1+\beta}} \left[ \frac{\Gamma}{p_{2}^{2}} \right]=
 \left(\frac{\sqrt{t}}{2 \sqrt{1+\beta}} \right)^{2/3}.
\end{equation}

\textbf{9. Comparison with the purely numerical calculation}

Some pictures will now be used to compare the uniform method with the results of
 numerical diagonalisation (which can be taken as the `exact' result). Figures
 \ref{fig:uniform0}--\ref{fig:uniform200} show a selection of Bloch waves with
 the uniform calculation shown as a solid line, though as before only the discrete
 values of $y$ corresponding to the diffracted beams were used. The dots are
 the numerical data.

\begin{figure}[htbp]
\begin{center}
\centerline{\epsfig{figure=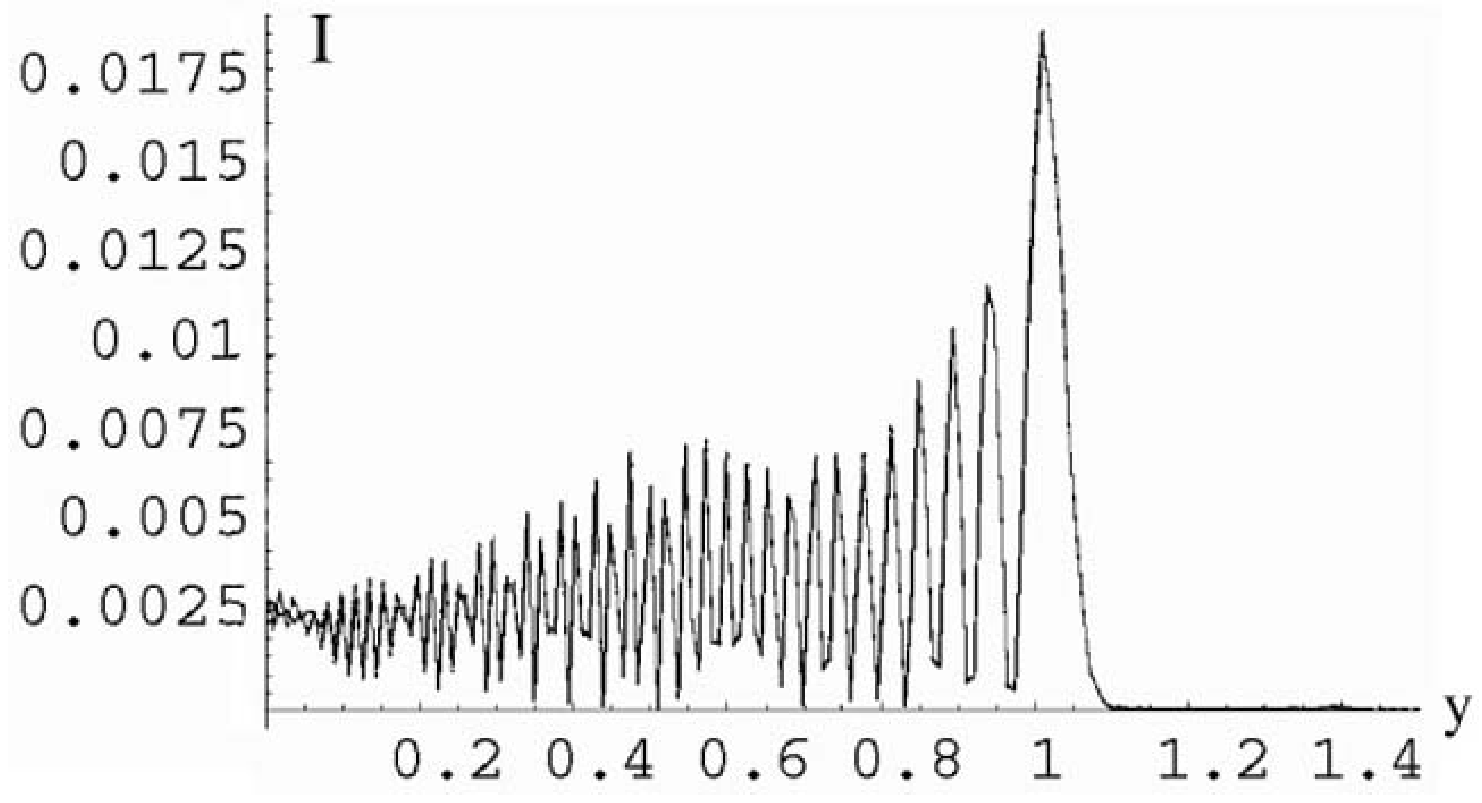,width=15cm,angle=0}}
\end{center}
\caption{A comparison of the farfield
wavefunction obtained by numerical diagonalisation (dashed), with the uniform 
calculation (solid), for $\Lambda=12500$ and $\zeta=\pi/2$.} 
\label{fig:unisuma}
\end{figure}

\begin{figure}[htbp]
\begin{center}
\centerline{\epsfig{figure=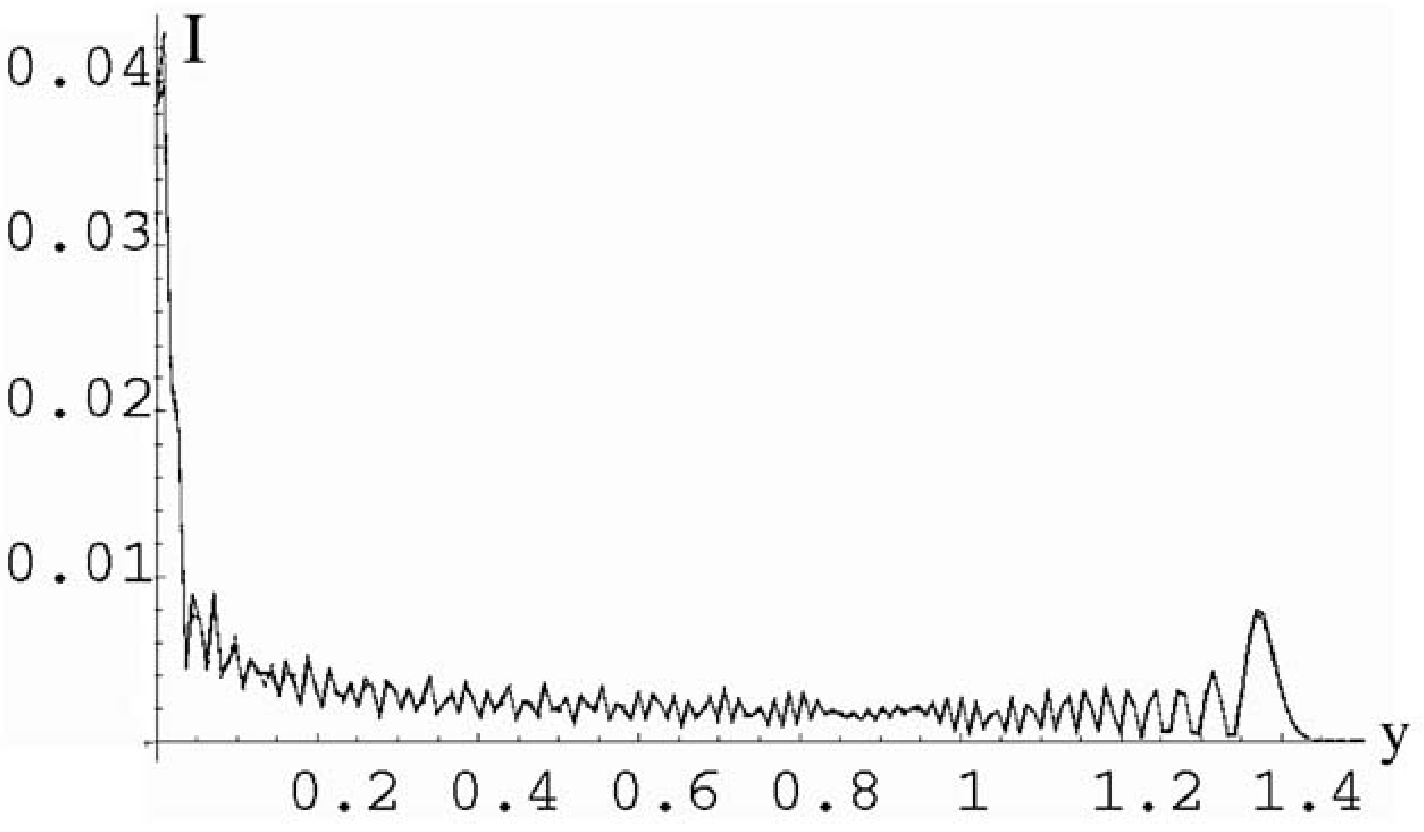,width=15cm,angle=0}}
\end{center}
\caption{A comparison of the farfield
wavefunction obtained by numerical diagonalisation (dashed), with the uniform 
calculation (solid), for $\Lambda=12500$ and $\zeta=\pi$.} 
\label{fig:unisumb}
\end{figure}

\begin{figure}[htbp]
\begin{center}
\centerline{\epsfig{figure=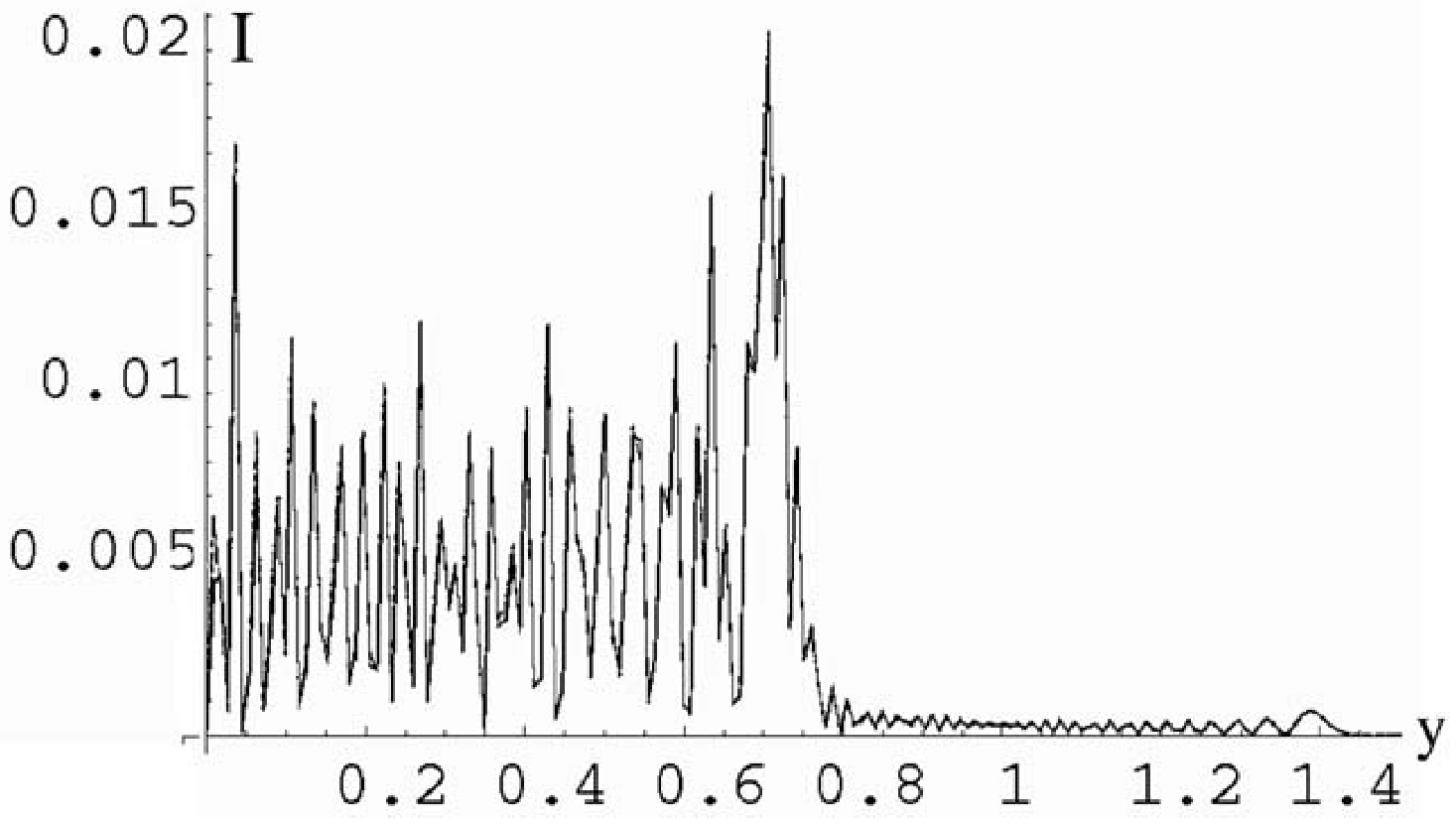,width=15cm,angle=0}}
\end{center}
\caption{A comparison of the 
farfield wavefunction obtained by numerical diagonalisation (dashed), with the 
uniform calculation (solid), for $\Lambda=12500$ and $\zeta=3\pi/2$.} 
\label{fig:unisum2a}
\end{figure}

\begin{figure}[htbp]
\begin{center}
\centerline{\epsfig{figure=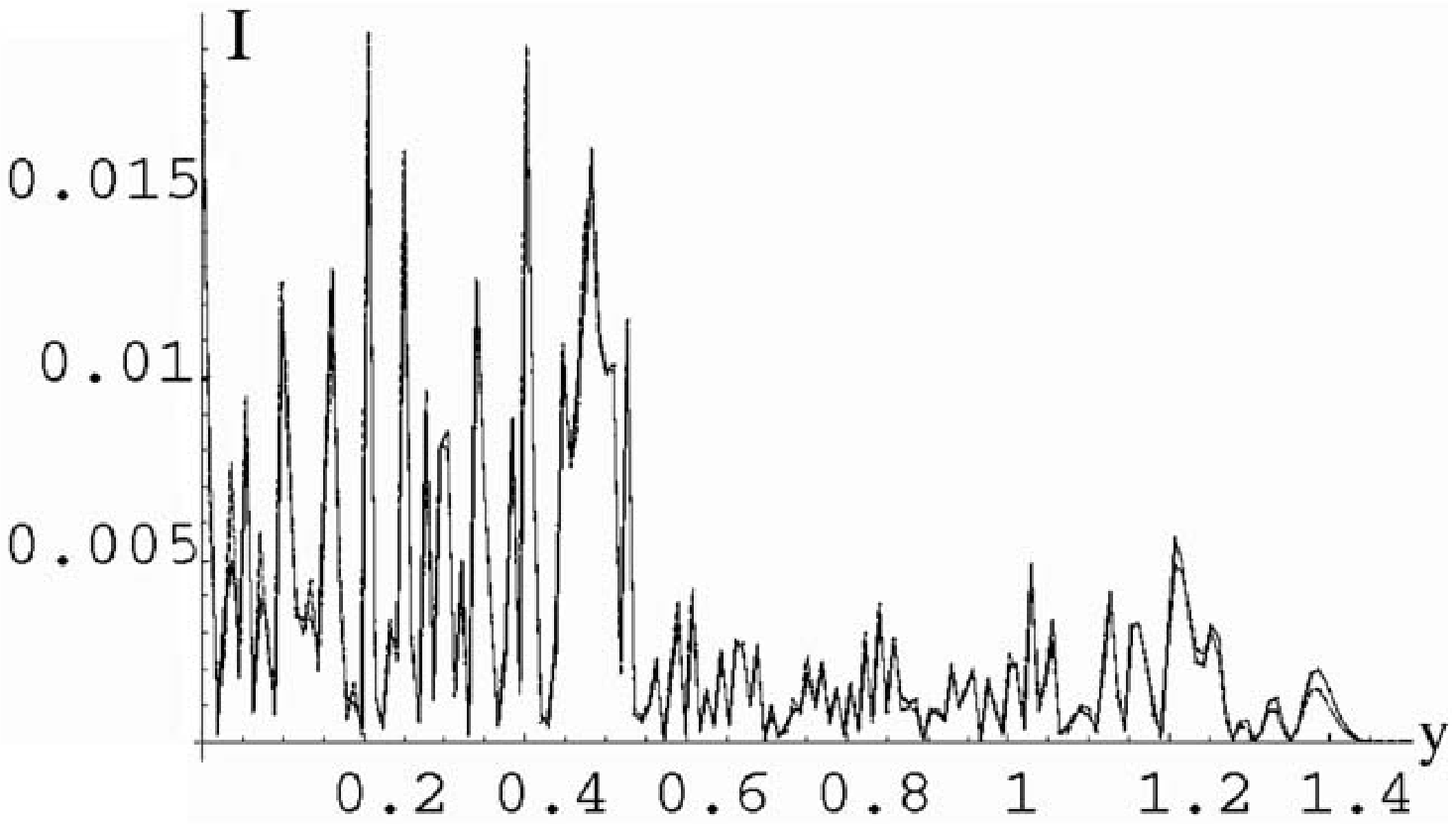,width=15cm,angle=0}}
\end{center}
\caption{A comparison of the 
farfield wavefunction obtained by numerical diagonalisation (dashed), with the 
uniform calculation (solid), for $\Lambda=12500$ and 
$\zeta=7\pi/2$.} \label{fig:unisum2b}
\end{figure}

\begin{figure}[htbp]
\begin{center}
\centerline{\epsfig{figure=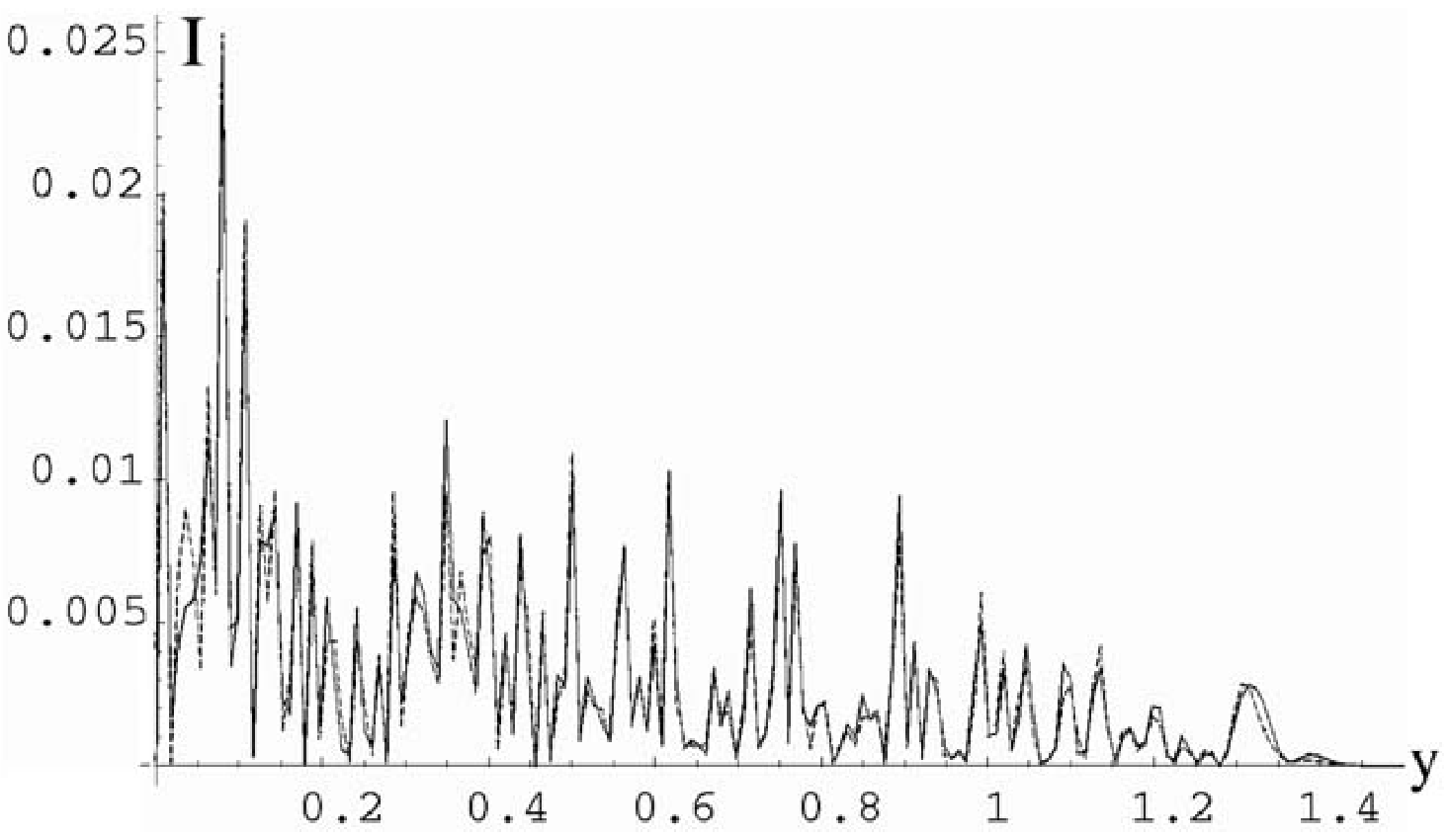,width=15cm,angle=0}}
\end{center}
\caption{A comparison of the farfield 
wavefunction obtained by numerical diagonalisation (dashed), with the uniform 
calculation (solid), for $\Lambda=12500$:  $\zeta=81 \pi/2$.} 
\label{fig:unisum3}
\end{figure}

Figures \ref{fig:unisuma}--\ref{fig:unisum3} give the square of the modulus 
of the total wavefunction
(i.e.\ the diffraction pattern),
found by summing the Bloch waves, for a selection of depths (i.e.\ thicknesses of
interaction region). As mentioned at the end 
of section 5, the superposition coefficients in this sum are given by 
the value 
of the particular Bloch wave at $y=0$. Only the bound states were used in both
 the numerical and uniform calculations, an approximation which is increasingly
accurate in the classical limit.
The intensity shown in Figure \ref{fig:unisuma} clearly displays (the square of)
an Airy function, which is well known to be the wavefunction
associated with a fold caustic (Berry 1981). Caustics 
are the foci of the diffracted wavefield---classically they diverge
and hence come to dominate the diffraction pattern in this limit. Figures
\ref{fig:unisuma}--\ref{fig:unisum3} are the quantum equivalent of
vertical slices through the classical ray trajectory
picture, Figure 1b, of Berry (1999) (the depth,
$\zeta$, is the same classical unit denoted by $x$ in that paper).
The caustics are the envelopes of families of rays (each ray corresponds 
to a classical atom) which oscillate back
and forth in the potential. 

If the potential was harmonic
then all the atoms would be focussed at $y=0$ when $\zeta=m \pi$ 
($m=0,1,2 \ldots$)
and this behaviour can be seen in Figure \ref{fig:unisumb} for $m=1$.
However, the anharmonicity of the sinusoidal potential means that
the foci are imperfect and are smeared out into cusps with fold caustic arms.
With increasing depth, successive oscillations of the atoms each introduce
a new cusp which develops into a fold as it moves outwards through the pattern
with increasing $\zeta$. The result is an increasingly complicated 
interference pattern between sucessive Airy functions as shown in Figures
\ref{fig:unisum2a}--\ref{fig:unisum3}, but usually with a few fringes of
an outlying Airy function visible. The proliferation of caustics for
increasing $\zeta$ is described only by dynamical diffraction and lies
beyond the phase grating (or so-called `Raman-Nath') approximation. 

\textbf{10. The problem of the separatrix}

As has been already been noted, when the eigenvalue $\beta$ approaches 1 the
 solution to the RN equation changes its nature. Classically,
the motion of a particle in the sinusoidal potential goes from being trapped
in a single well (libration) to being free (rotation) when its energy passes
 through the separatrix at $\beta=1$ from below. At the same time the number
 of turning points of the
`momentum' function $p_{1}$ jumps from two to four and in so doing $p_{1}$
 becomes qualitatively different from $p_{2}$---as apparant from 
Figure \ref{fig:two-momenta} when the central dip breaks through the zero line.
These new turning-points occur at 
\begin{equation}
y=\pm \sqrt{\beta-1}. 
\end{equation} 
For small values of $\Lambda$ the actual divergences due to these turning-points
 can fall between the diffracted orders and go unnoticed. As $\Lambda$ is
 increased this is no longer the case and the divergences become clearly defined
 as the classical distribution emerges. It is emphasised that these divergences
 only affect those eigenvectors with eigenvalues close to the separatrix. Careful
 examination of the
 picture of the $200^{\mathrm{th}}$ eigenstate for $\Lambda=12500$, reveals the
 first hint that the uniform approximation has a defect when $\beta$ begins to
 approach 1. The remainder of this paper is concerned with these eigenstates close
 to the separatrix. Although only forming a small fraction of the total eigenstate
 sum giving the diffracted wavefunction, they are perhaps the most interesting states
 as they contain the very fine corrections due to tunnelling between the coordinate
 space potential wells
in the semiclassical limit.

\begin{figure}[htbp]
\begin{center}
\centerline{\epsfig{figure=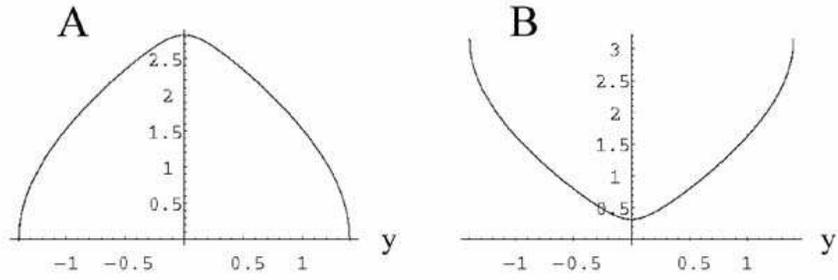,width=12cm,angle=0}}
\end{center}
\caption{A: The original $p_{2}$, and B:
transformed $\bar{p}_{2}$, phase momenta for $\beta=0.95$.}
\label{fig:separatrixmom}
\end{figure}

The uniform approximation used so far was not designed to handle the new
turning-points. The momentum function $p_{2}$ used in the mapping relation
 (\ref{eq:mapping-function}) contains no information concerning the new 
turning-points. The amplitude and phase functions no longer act in concert.
 One way to proceed is by a transformation upon the RN equation which results in
 both the momentum functions have turning-points at $y=\sqrt{\beta-1}$.
Defining
\begin{equation}
B_{n}=(-1)^{n}C_{n} \label{eq:separatrix-transformation} 
\end{equation}
the stationary R-N equation (\ref{eq:rn-stationary-scaled}) becomes
\begin{equation}
(y^{2}-\beta)C_{n}+\frac{1}{2}(C_{n+1}+C_{n-1})=0. \label{eq:rn-stationary-separatrix}
\end{equation}
from which one obtains an altered action 
\begin{equation}
\frac{\partial S}{\partial y}=\sqrt{\Lambda}\;  \arccos \; [\beta-y^{2}] +
\mathrm{i}\frac{(\beta-y^{2})y}{1-(\beta-y^{2})^{2}}
\end{equation}
leading to the WKB formula
\begin{equation}
C(y)=\frac{\mathrm{e}^{\pm \mathrm{i}\sqrt{\Lambda}\; \int  \arccos \; 
[\beta-y^{2}] \; dy}}{\left(1-(y^{2}-\beta)^{2} \right)^{1/4}} =  
\frac{\mathrm{e}^{ \pm \mathrm{i}\sqrt{\Lambda}\; \bar{S}_{0}(y,\beta)}}
{\left(1-(y^{2}-\beta)^{2} \right)^{1/4}}.
\label{eq:wkb-separatrix}
\end{equation}
The amplitude is the same as before, giving the two sets of turning-points,
 but the phase momentum
\begin{equation}
\bar{p}_{2}(y,\beta)= \arccos \; [ \beta - y^{2}]
\end{equation}
now has its turning-points at $y=\pm \sqrt{\beta-1}$ as promised. The cost is the
 loss of the turning-points at $y= \pm \sqrt{\beta+1}$. The effect of the transformation
 is in swapping the r\^{o}les of inner and outer turning-points. Figure 
\ref{fig:separatrixmom} shows that whereas $p_{2}$ has the momentum profile 
for a well, $\bar{p}_{2}$ has that of barrier which the particle has enough 
energy to surmount. 

Since the phase momentum functions $p_{2}$ and $\bar{p}_{2}$ only describe 
one set of turning-points each, one is forced into employing two separate 
transitional uniform approximations for each eigenfunction when one is close 
to the separatrix. One transitional approximation covers the inner turning 
points and the other the outer. Both are valid in the intermediate region where 
they smoothly join.

\textbf{11. The parabolic barrier equation}

The inner turning-points require a transitional approximation for a (smooth)
 potential barrier. A suitable comparison equation is
\begin{equation}
\frac{d^{2} \phi}{d \sigma^{2}}+\left(t+\sigma^{2}\right) \phi = 0. 
\label{eq:barrier-eqn}
\end{equation}
For $t>0$, the Bloch states are more energetic than the central potential 
barrier, and classically one has transmission above the barrier. This is 
referred to as the \emph{underdense} case.
Making the change of variables
\begin{eqnarray}
t & = & \mathrm{i} \bar{t} \\
\sigma & = &  \frac{ \bar{\sigma}}{\sqrt{2}} \mathrm{e}^{\mathrm{i} \pi/4}
\end{eqnarray}
one is lead back to the equation 
\begin{equation}
\frac{d^{2} \phi}{d \bar{\sigma}^{2}}-\left(\frac{\bar{t}}{2}+
\frac{\bar{\sigma}^{2}}{4}\right) \phi = 0 \label{eq:underdense-barrier-eqn}
\end{equation}
which is the same as the first (upper sign) parabolic cylinder equation 
(\ref{eq:parabolic-cylinder-equations}) when the identifications 
$g  =  \bar{\sigma}$ and  $a  = \bar{t}/2$ are made.

The appropriate solutions for the barrier top are not the Whitakker functions 
since a parabolic barrier does not give an exponentially decaying wavefunction 
for large $g$. Instead, for a barrier, the basic even and odd power series solutions, 
which will be referred to as $\Theta_{1}(a,g)$ and $\Theta_{2}(a,g)$ respectively, 
are the correct choice. Close to the barrier top $a$ is small and the power series 
solutions are most conveniently expressed in terms of the confluent hypergeometric 
functions  
\begin{eqnarray}
\Theta_{1}(a,g) & = &\mathrm{e}^{-g^{2}/4} \hspace{1em}_{1} \! F_{1} 
\left(\frac{a}{2}+\frac{1}{4};\frac{1}{2};\frac{g^{2}}{2} \right) \\
\Theta_{2}(a,g) & = & g \mathrm{e}^{-g^{2}/4} \hspace{1em}_{1} \! F_{1} 
\left(\frac{a}{2}+\frac{3}{4};\frac{3}{2};\frac{g^{2}}{2} \right).
\end{eqnarray}
As discussed previously, the boundary conditions mean that only even eigenfunctions
 are of interest here. Thus, the underdense inner turning point transitional
 approximation will be based upon the even power series 
\begin{equation}
\Theta_{1}\left(-\mathrm{i} \frac{t}{2}, \sqrt{2} \sigma 
\mathrm{e}^{-\mathrm{i} \pi/4} \right).
\end{equation}

\textbf{12. The action for an underdense barrier}

The underdense barrier does not induce any real turning-points (though of 
course the proximity of the turning-points to the real axis gives the deviation 
of the WKB amplitude from the true value) so the natural choice of reference 
point from which to integrate the phase is $y=\sigma=0$. One finds
\begin{equation}
\begin{split}
\bar{S}_{0}(0,y,\beta)& =\int_{0}^{y} \arccos \left[\beta-{y'}^{2} \right] \; 
dy' \\
& = y \arccos \left[\beta - y^{2} \right] + 2 \mathrm{i} \sqrt{1- \beta} \; 
\mathrm{E} \left. \left(\frac{1}{2} \arccos \left[\beta-y^{2} \right] \right| 
\frac{2}{1 - \beta} \right) \\ 
& \qquad \qquad -\mathrm{E} \left. \left(\frac{1}{2} \arccos \left[\beta \right] 
\right| \frac{2}{1 - \beta} \right).
\end{split}
\end{equation}
Although it appears that this action contains an imaginary piece this is 
actually not the case. Strictly, the well known transformations (see 
Abromowitz and Stegun 1964) 
should be applied to the elliptic functions so that their parameters lie between 
zero and one (the parameter used above tends to infinity as $\beta \rightarrow 1$). 
When this is done the action $\bar{S}_{0}$ is explicitly real. However, the 
transformations produce more complicated expressions so will not be applied here. 

To find the value of $t$, which was previously given by the integral across the well, 
one must now integrate up the imaginary axis between the points
\begin{equation}
y_{\pm}= \pm \sqrt{\beta-1}= \pm \mathrm{i} \sqrt{1-\beta}.
\end{equation}
The equivalent points for the underdense barrier comparison equation 
(\ref{eq:barrier-eqn}) are
\begin{equation}
\sigma= \pm \mathrm{i} \sqrt{t}.
\end{equation}
Letting $y=\mathrm{i}v$ and $\sigma=\mathrm{i} \varsigma$, $t$ is implicitly 
given by
\begin{equation}
2\mathrm{i}\sqrt{\Lambda}\int_{0}^{\sqrt{1- \beta}} \arccos \left[ \beta + 
v^{2} \right] \;dv = 2 \mathrm{i} \int_{0}^{\sqrt{t}} \sqrt{t^{2}-\varsigma^{2}} 
\;d \varsigma
\end{equation}
which, in a similar to fashion to before, results in the condition
\begin{equation}
4 \sqrt{\Lambda} \sqrt{1- \beta} \mathrm{E} \left. \left(\frac{1}{2} \arccos 
\; [\beta] \right| \frac{2}{1-\beta} \right)= \frac{t \pi}{2}. 
\label{eq:underdense-t}
\end{equation}
The mapping function between $\sigma$ and $y$ implicitly giving $\sigma(y)$ is
\begin{equation}
\sqrt{\Lambda} \bar{S}_{0}(0,y,\beta)=\int_{0}^{\sigma} \sqrt{t+\sigma^{2}} \; 
d \sigma = \frac{t}{2}\left( \mathrm{arccosh}\left[\sqrt{1+\frac{\sigma^{2}}{t}} 
\right] + \frac{\sigma}{\sqrt{t}} \sqrt{1+ \frac{\sigma^{2}}{t}} \; \right) 
\label{eq:underdense-mapping-relation}
\end{equation}
which, together with the value of $t$, gives the transitional approximation to 
the wavefunction
\begin{equation}
\begin{split}
B(y)=(-1)^{n}C(y) & =(-1)^{n} \psi_{\mathrm{transitional}} \\ & \propto \quad 
(-1)^{n} \left( \frac{t+ \sigma^{2}}{1-\left(y^{2}-\beta \right)^{2}} \right)^{1/4} 
\mathrm{e}^{\mathrm{i} \sigma^{2}/2} \; \hspace{1em}_{1} \! F_{1} 
\left(-\mathrm{i}\frac{t}{4}+\frac{1}{4}; \frac{1}{2}; -\mathrm{i} 
\sigma^{2} \right). 
\label{eq:transitional-wavefunction}
\end{split}
\end{equation}
This is a real function for real $t$ and $\sigma$, which is valid from $y=0$ 
and almost all the way to the outer turning-point, breaking down close to it 
because it is only set up to deal with the inner turning-point. The constant of 
proportionality will now be obtained by matching the asymptotic behaviour of 
this function to the WKB solution somewhere between the two transition points.
 
\textbf{13. The asymptotics of the barrier transitional approximation}

\begin{figure}[htbp]
\begin{center}
\centerline{\epsfig{figure=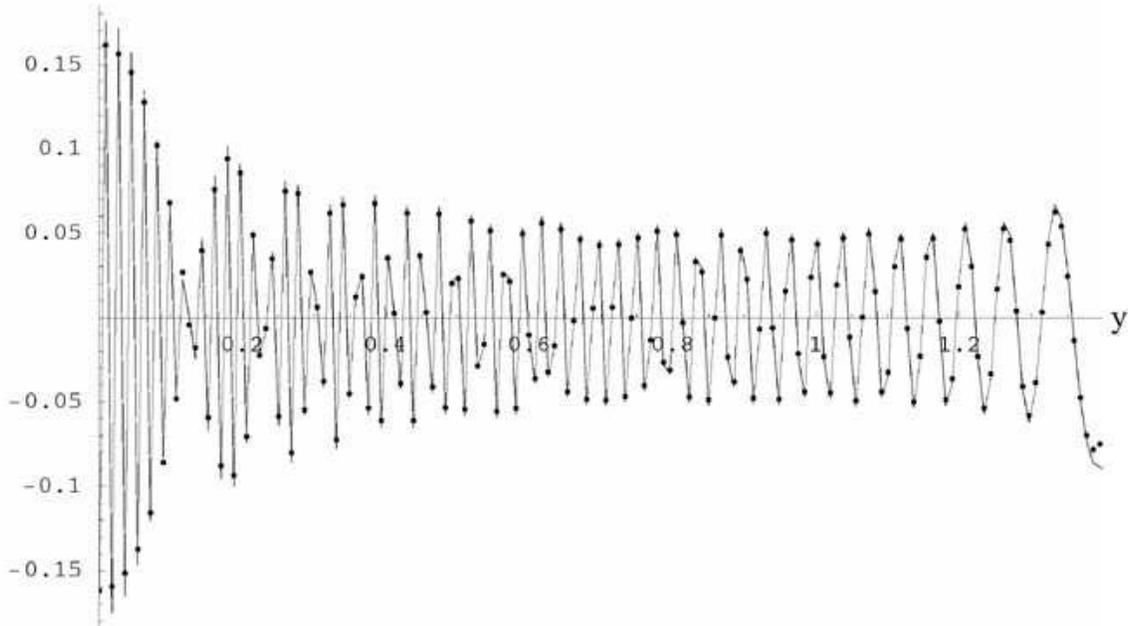,height=15cm,angle=-90}}
\end{center}
\caption{The WKB approximation using the 
underdense parabolic barrier action, see Equation (\ref{eq:wkb-with-mu}), and 
the fully numerical solution. In particular, this tests the derived 
phase angle 
$\mu$ as given by (\ref{eq:mu-parabolic}). The dots are the numerically 
calculated points, and the continuous line joins the WKB amplitudes. The 
value of $\Lambda$ is 12500 and $\beta=0.9961$.} 
\label{fig:wkb-underdense}
\end{figure}

The confluent hypergeometric function has well known asymptotics. When $|\sigma|$ is large
\begin{equation}
\begin{split}
\hspace{1em}_{1} \! F_{1} \biggl(-\mathrm{i}\frac{t}{4}+ & \frac{1}{4}; \frac{1}{2}; 
-\mathrm{i} \sigma^{2} \biggr)  \\ & = \frac{\Gamma \left(\frac{1}{2} \right)}{\Gamma 
\left( \frac{1}{4} + \mathrm{i} \frac{t}{4} \right)}  \mathrm{e}^{-\mathrm{i}\pi 
( -\mathrm{i}t+1 )/4} \left(-\mathrm{i} \sigma^{2} \right)^{(\mathrm{i}t-1)/4} 
\left(1+ \frac{(\mathrm{i} t^{2}-4t-3\mathrm{i})}{16 \sigma^{2}} + 
\mathcal{O}\left(\frac{1}{\sigma^{4}} \right) \right) \\ & \qquad + 
\frac{\Gamma \left(\frac{1}{2} \right)}{\Gamma \left( \frac{1}{4} - 
\mathrm{i} \frac{t}{4} \right)} \mathrm{e}^{-\mathrm{i}\sigma^{2}} 
\left(-\mathrm{i} \sigma^{2} \right)^{-(\mathrm{i}t+1)/4} \left(1+ 
\frac{(-\mathrm{i} t^{2}-4t+3\mathrm{i})}{16 \sigma^{2}} + 
\mathcal{O}\left(\frac{1}{\sigma^{4}} \right) \right)
\end{split}
\end{equation}
where $\Gamma$ is the Gamma (factorial) function. So
\begin{equation}
\begin{split}
\Theta_{1} \biggl(- & \mathrm{i}\frac{t}{2}, \sqrt{2} \sigma \mathrm{e}^{-\mathrm{i}
 \pi/4} \biggr) =
\mathrm{e}^{\mathrm{i} \sigma^{2}/2} \; \hspace{1em}_{1} \! F_{1} 
\left(-\mathrm{i}\frac{t}{4}+\frac{1}{4}; \frac{1}{2}; -\mathrm{i} \sigma^{2} \right) \\ 
& \sim \Gamma \left( \frac{1}{2} \right)  \left(-\mathrm{i} \sigma^{2} \right)^{-1/4} 
\biggl(  \mathrm{e}^{-\pi(t+\mathrm{i})/4} \mathrm{e}^{\mathrm{i} \sigma^{2}/2} 
\frac{\left( -\mathrm{i} \sigma^{2} \right)^{\mathrm{i}t/4}}{\Gamma 
\left(\frac{1}{4}+\mathrm{i}\frac{t}{4} \right)} \mathrm{e}^{\ln 
\left[1+(\mathrm{i} t^{2}-4t-3\mathrm{i})/(16 \sigma^{2})\right]} \\
& \qquad \qquad + \mathrm{e}^{-\mathrm{i} \sigma^{2}/2} 
\frac{\left( -\mathrm{i} \sigma^{2} \right)^{-\mathrm{i}t/4}}{\Gamma 
\left(\frac{1}{4}-\mathrm{i}\frac{t}{4} \right)} \mathrm{e}^{\ln 
\left[ 1-(\mathrm{i} t^{2}+4t-3\mathrm{i})/(16 \sigma^{2})\right]} \biggr)
\end{split}
\end{equation}
which conveniently reduces to
\begin{equation}
\begin{split}
\Theta_{1} & \biggl( - \mathrm{i}\frac{t}{2}, \sqrt{2} \sigma \mathrm{e}^{-\mathrm{i} 
\pi/4} \biggr) \\ & \sim 2 \frac{\Gamma \left( \frac{1}{2} \right)}{\left| \Gamma 
\left(\frac{1}{4}+ \mathrm{i} \frac{t}{4} \right) \right|} \sigma^{-1/2} 
\mathrm{e}^{-\pi t/8-t/(4 \sigma^{2})} \cos \left( \frac{t}{2} \ln \sigma + 
\frac{ \sigma^{2}}{2} -\mathrm{Arg} \left[ \Gamma \left(\frac{1}{4} + \mathrm{i} 
\frac{t}{4} \right) \right] -\frac{\pi}{8}+ \mathcal{O} \left(\frac{1}{\sigma^{4}} 
\right) \right). \label{eq:1F1-asymptotic}
\end{split}
\end{equation}
When $\Lambda$ is large enough $\sigma$ quickly takes on large values for even 
modest sizes of $y$, and so the confluent hypergeometric function attains its 
asymptotic form in the region between the inner and outer turning-points. It 
may then be compared to the WKB solution (\ref{eq:wkb-separatrix}) for the transformed 
RN equation. The left and right travelling WKB waves (\ref{eq:wkb-separatrix}) are 
combined to give a real solution
\begin{equation}
B_{\mathrm{barrier}}(y)=\frac{\mathcal{N}}{\left(1-\left(y^{2}- \beta \right)^{2} 
\right)^{1/4}} \cos \left( \sqrt{\Lambda} \bar{S}_{0}(0,y, \beta) + \mu(\beta) + 
\sqrt{\Lambda} \pi y \right) \label{eq:wkb-with-mu}
\end{equation}
where the $(-1)^{n}$ factor has been incorporated into the phase of the cosine as 
$\sqrt{\Lambda} \pi y$. To enable a direct comparison, the phase of the cosine of 
Equation (\ref{eq:1F1-asymptotic}) should also be augmented by the same quantity.
The real phase angle $\mu(\beta)$ for this parabolic barrier approximation (which 
for a simple first order turning-point, due to a linear potential, is equal to 
$\pi/4$) will this time be determined by consistency with the asymptotic solution 
(\ref{eq:1F1-asymptotic}). 
In order for a comparison to be made, the action $\bar{S}_{0}(0,y,\beta)$ appearing 
in the WKB solution must be written in terms of $(\sigma,t)$, which is accomplished 
through Equation (\ref{eq:underdense-mapping-relation}). Expanding the rhs of 
(\ref{eq:underdense-mapping-relation}) for $\sigma \gg t$, one has
\begin{equation} 
\frac{t}{2}\left( \mathrm{arccosh}\left[\sqrt{1+\frac{\sigma^{2}}{t}} \right] + 
\frac{\sigma}{\sqrt{t}} \sqrt{1+ \frac{\sigma^{2}}{t}} \; \right) \sim \frac{t}{2} 
\ln \sigma - \frac{t}{4} \ln t +\frac{t}{2} \ln 2+ \frac{\sigma^{2}}{2} + 
\frac{t}{4} + \mathcal{O}\left(\frac{t^{2}}{\sigma^{2}} \right)
\end{equation}
implying that
\begin{equation}
\mu=\frac{t}{4} \ln t -\frac{t}{2} \ln 2 -\frac{t}{4}- \mathrm{Arg} 
\left[ \Gamma \left(\frac{1}{4}+ \mathrm{i} \frac{t}{4} \right) \right]- 
\frac{\pi}{8}. \label{eq:mu-parabolic}
\end{equation}
Figure \ref{fig:wkb-underdense} demonstrates that this expression for $\mu$ is
 correct by comparing the WKB solution (\ref{eq:wkb-with-mu}) containing it, 
with the fully numerical calculation. The value of $\Lambda$ is reasonably 
small so the WKB solution diverges only very slightly from the correct value.

The exact solution to the parabolic cylinder equation has thus contributed to the 
evaluation of the phase of the WKB solution. On the other hand, the WKB solution 
indicates the necessary modifications needed for the amplitude of the parabolic 
cylinder equation so that it becomes the correct transitional solution to the 
particular problem being dealt with. Equating the ampliutdes of Equations 
(\ref{eq:1F1-asymptotic}) and (\ref{eq:wkb-with-mu}), one finds Equation 
(\ref{eq:transitional-wavefunction}) can now be updated to read
\begin{equation}
B_{\mathrm{barrier}}(y)=(-1)^{n} \mathcal{N} \frac{\left| \Gamma 
\left(\frac{1}{4}+ \mathrm{i} \frac{t}{4} \right) \right| 
\mathrm{e}^{\pi t/8}}{2 \Gamma \left( \frac{1}{2} \right)} 
\left( \frac{t+ \sigma^{2}}{1-\left(y^{2}-\beta \right)^{2}} 
\right)^{1/4} \mathrm{e}^{\mathrm{i} \sigma^{2}/2} \; \hspace{1em}_{1} 
\! F_{1} \left(-\mathrm{i}\frac{t}{4}+\frac{1}{4}; \frac{1}{2}; 
-\mathrm{i} \sigma^{2}\right). \label{eq:underdense-wavefunction}
\end{equation}

\begin{figure}[htbp]
\begin{center}
\centerline{\epsfig{figure=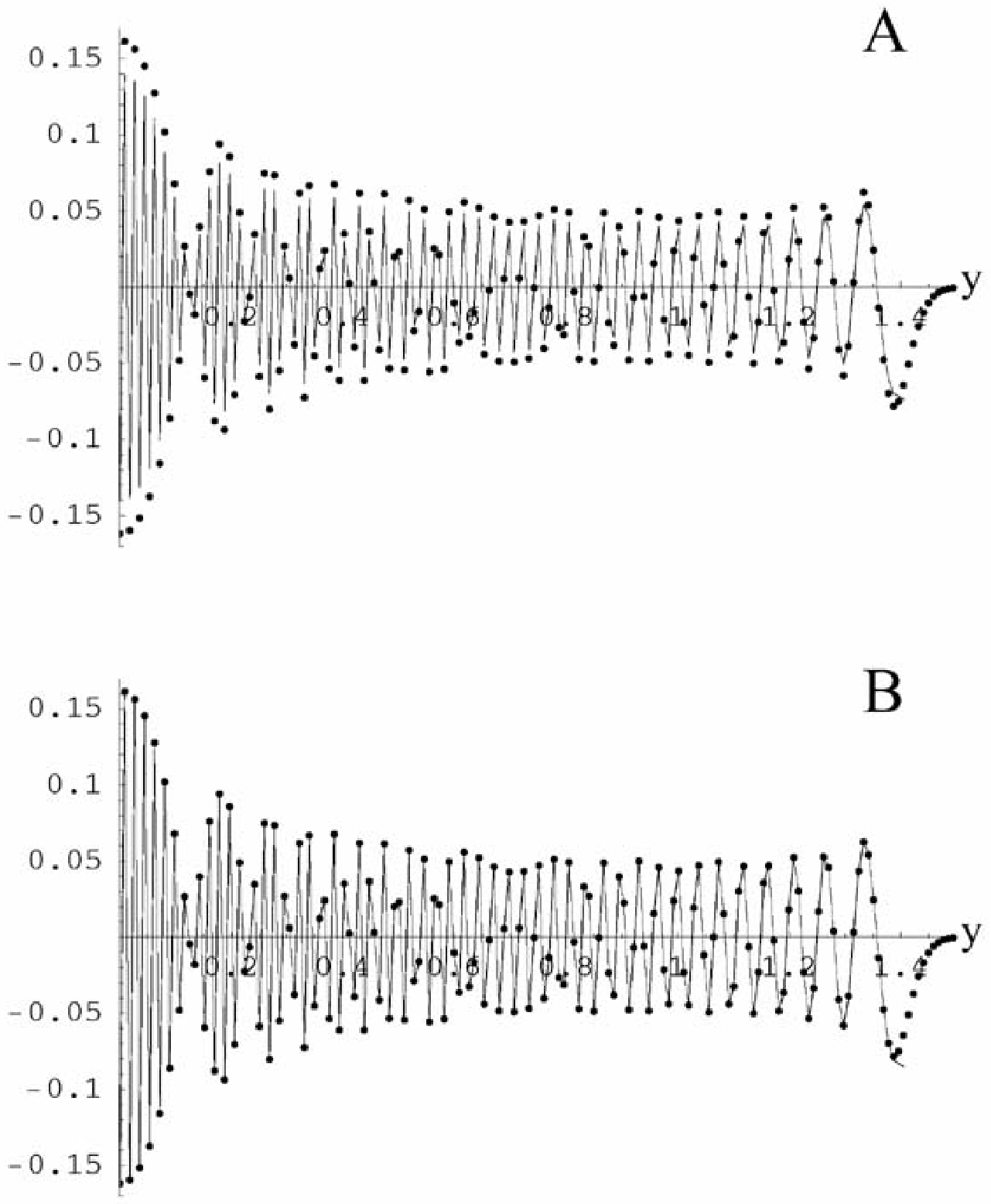,width=15cm,angle=0}}
\end{center}
\caption{The parabolic barrier transitional 
approximation: A) as given by Equation (\ref{eq:underdense-wavefunction}); B) 
the renormalised version. The dots are the fully numerical solution. The value 
of $\Lambda$ is 12500 and $\beta=0.9961$.}
\label{fig:trans-under}
\end{figure}

The parabolic transitional approximation is compared to the fully numerical 
result in Figure \ref{fig:trans-under} A). 
At first sight the match does not seem too good. The reason is that the normalisation 
uses  the WKB amplitude factor, which diverges at the turning-points. When there are 
only the outer turning-points this method seems to work (see Figures 
\ref{fig:uniform0}-\ref{fig:uniform152}) since the divergences are narrow enough to 
not produce too significant a contribution. However, the appearance of the inner 
turning-point divergences close to the separatrix energy now means the normalisation 
factor is significantly over estimating the magnitude of the wavefunction, and thus 
reduces the magnitude too much as shown. With relatively little effort one can 
numerically normalise the uniformly calculated eigenvectors by summing the discrete 
amplitudes, and when this is carried out the match, shown in Figure 
\ref{fig:trans-under} B), is exceedingly good. This illustrates that it is 
only the normalising factor which is at fault.
Figure \ref{fig:trans-under} B) further illustrates that the barrier 
transitional approximation, Equation (\ref{eq:underdense-wavefunction}), 
is correct nearly throughout the entire momentum range---only breaking down 
close to the \emph{outer} turning-point.

\textbf{14. Calculation of the eigenvalues close to the separatrix---a 
modified Bohr-Sommerfeld rule}

There is a slight complication to the calculation of the allowed values of 
$\beta$ close to the separatrix which needs to be highlighted. When comparing 
the values of $\beta$ obtained by the numerical diagonalisation technique with 
those obtained via Equation (\ref{eq:bohr-sommerfeld}), the two differ when $\beta$ 
grows very close to one.  Somehow the derivation of the basic WKB solution 
(\ref{eq:wkb}) has failed to capture the full behaviour of the $p_{2}$ 
function---perhaps it should now afterall contain two turning-points, 
not one, and so match the structure of the amplitude $p_{1}$ term? 
(Implying the transformation (\ref{eq:separatrix-transformation}) of the 
phase momentum is more than a device.) From the point of view of the 
eigenvectors this can be overcome by replacing the previous single uniform 
approximation with two transitional approximations when $\beta$ approaches 
one; the parabolic transitional approximation to cover the inner turning-point, 
and an Airy function approximation for the outer turning-point (since this 
remains a simple first order turning-point).
However, to calculate the allowed values of the action which corresponds to the 
bound states, one needs some expression which is valid throughout the entire 
region which joins the two turning-points.

The general procedure for finding the action across a classically allowed 
region which separates two arbitrary types of turning-point employs two 
transitional approximations which are each valid at one end of the region, 
but these must be correctly joined. The quantised values of $S$, and hence 
$\beta$, are those which correctly match the two somewhere in the region of 
mutual validity.  

The matching is most easily accomplished using the asymptotic forms for the two 
transitional approximations---which are of course their WKB approximations. 
In the region between the two turning-points one thus has
\begin{equation}
\begin{split}
\frac{1}{\left(1-\left(y^{2}-\beta \right)^{2} \right)^{1/4}} & \cos 
\left( \sqrt{\Lambda} \bar{S}_{0}(0,y,\beta)+ \mu(\beta)+\sqrt{\Lambda} 
\pi y \right) \\ & =\frac{1}{\left(1-\left(y^{2}-\beta \right)^{2} \right)^{1/4}} 
\cos \left(\sqrt{\Lambda} S_{0}(\sqrt{1+\beta},y,\beta)+ \frac{\pi}{4} \right) 
\label{eq:cos-cos-match}
\end{split}
\end{equation}
which implies that
\begin{equation}
\sqrt{\Lambda} \bar{S}_{0}(0,y,\beta)+ \mu(\beta)+\sqrt{\Lambda} \pi 
y=\sqrt{\Lambda} S_{0}(\sqrt{1+\beta},y,\beta)+ \frac{\pi}{4}
\end{equation}
modulo $2 \pi$.

The method described above works in conventional situations with WKB expressions 
developed from (continuous) differential equations. Once again however, the approach 
has to be modified for the RN equation---whilst successful for the single well, as 
soon as the inner turning-points begin to approach the real axis even the matching 
of the two transitional approximations runs into trouble. The reason is that the 
continuous descriptions embodied above by Equation (\ref{eq:cos-cos-match}) do not 
match at all. Only when they are evaluated at the discrete points corresponding to 
diffracted beams do they match. The transformation (\ref{eq:separatrix-transformation}) 
has produced two different equations whose continuised WKB expressions only respect 
their common origin at the discrete level.
It is then a surprise to find that at the correct (characteristic) values of $\beta$ 
the discretely evaluated expressions on either side of Equation (\ref{eq:cos-cos-match}) 
are in perfect agreement for all $y$. Both are identical in each other's supposedly 
exclusive region of validity. This is rather curious, but the characteristic values 
of $\beta$, which one is able to predict by correctly matching the discrete points 
of the two WKB expressions, demonstrate that it is correct. 

Due to the simultaneous validity for all $y$, the most sensible point to choose to 
match the two solutions is $y=0$. The correct matching condition for even eigenstates 
becomes one of 
\begin{eqnarray}
\cos \left(\mu(\beta) \right) - \cos \left( \sqrt{\Lambda} S_{0}(y_{+},0,\beta) + 
\pi/4 \right) & = & 0 \label{eq:underdense-eigenvalue-condition-up}\\
\cos \left(\mu(\beta) \right) - \cos \left( \sqrt{\Lambda} S_{0}(y_{+},0,\beta) + 
5 \pi/4 \right) & = & 0 \label{eq:underdense-eigenvalue-condition-down}
\end{eqnarray}
the choice depending on whether the terminating Airy function has its peak above 
or below the $y$ axis. In fact, successive even eigenstates alternate between the 
two conditions. 
When using (\ref{eq:underdense-eigenvalue-condition-up}) and 
(\ref{eq:underdense-eigenvalue-condition-down}), it is necessary to express 
$\mu$, which is in the first instance a function of $t$, see Equation 
(\ref{eq:mu-parabolic}), as a function of $\beta$ through the definition of $t$ 
(Equation (\ref{eq:underdense-t})).  A further subtlety concerning the use of  
(\ref{eq:underdense-eigenvalue-condition-up}) and 
(\ref{eq:underdense-eigenvalue-condition-down}) is that close to each of the 
characteristic values there is another zero which does not correspond to an 
eigenvalue. The correct zeros are those through which the l.h.s.\ of Equations 
(\ref{eq:underdense-eigenvalue-condition-up}) and 
(\ref{eq:underdense-eigenvalue-condition-down}) have negative gradients. Table 
\ref{tab:evalues} compares the values of the top eight bound eigenvalues for 
$\Lambda=12500$ as calculated by the different methods which have been outlined so far. 
Clearly the modified method gives excellent agreement with the true value and is superior 
to the regular Bohr-Sommerfeld scheme when close to the separatrix. The remaining error 
between the modified method and the true value becomes smaller as $\Lambda \rightarrow 
\infty$. This is further emphasised by the last entry on the table which is the last 
bound eigenvalue for $\Lambda=250000$. The Bohr-Sommerfeld method predicts only 898 
even bound states whereas the modified method accurately finds the value of the 
$900^{\mathrm{th}}$.
 The subtle behaviour of the eigenvalues near the separatrix
can be physically attributed to the (semiclassically) exponentially small corrections 
due to wavefunction tunnelling. This phenomena has been recently discussed in a 
similar context by Waalkens, Wiersig and Dullin (1997) and by Sieber (1997).

\textbf{15. The Airy transitional approximation}

As has already been pointed out, to obtain the complete wavefunction correct for all 
$y$ one must join the parabolic barrier approximation (\ref{eq:underdense-wavefunction}) 
to another transitional approximation which covers the outer, first order, turning-point.
The comparison equation is given as an example in the Appendix 
(Equation (\ref{eq:airy-comparison-equation})), and choosing the reference 
point as $y=y_{+}=\sqrt{1+\beta}$, the mapping function $\sigma (y)$ is given by
\begin{equation}
 \sqrt{\Lambda} S_{0}(y_{+},y,\beta)
=\left\{ \begin{array}{ll}  - \frac{2}{3} \left|\sigma \right|^{3/2} & \mbox{if $y 
\leq \sqrt{1+\beta} \ ; \qquad (\sigma<0)$} \\[2ex]  \frac{2}{3} \mathrm{i} \sigma^{3/2} 
& \mbox{if $y > \sqrt{1+\beta} \ ; \qquad (\sigma>0)$} \end{array} \right.
\end{equation}
since the expression given for $S_{0}$, Equation (\ref{eq:wkb-phase}), is positive 
imaginary when $y>y_{+}$, and negative real when $y<y_{+}$.

The well known asymptotics of $\mathrm{Ai}(\sigma)$ when $\sigma \gg 0$ are 
\begin{equation}
\mathrm{Ai} (\sigma) \sim \frac{1}{2 \pi} \;  \sigma^{-1/4} \; 
\mathrm{e}^{-\frac{2}{3} \sigma^{3/2}}
\end{equation}
and so the Airy transitional approximation becomes
\begin{equation}
\psi_{\mathrm{transitional}}=B_{\mathrm{Airy}}(y)=2 \pi \mathcal{N} \left( 
\frac{\sigma (y)}{1- \left(y^{2}- \beta \right)^{2}} \right)^{1/4} \mathrm{Ai} 
\left(\sigma (y) \right).
\end{equation}

\textbf{16. The free eigenstates}

As emphasised previously, `free' is a description which refers to the (actual) 
configuration space situation of states having transverse energies greater than 
$V_{0}$. In (actual) momentum space there are no free states, the classical 
bounding of the maximum being set by the initial transverse momentum plus whatever 
the atoms can extract from the potential---which depends on the (actual) configuration 
space point, but has a maximum of $\sqrt{2mV_{0}}$. Thus,  even for $\beta>1$, one 
expects caustics in (actual) momentum space. One sees  why the free eigenstates 
are quantised and not continuous in energy. Somewhat perversely, the states which 
are free in (actual) configuration space, sit in a double well in (actual) momentum 
space, and so the central barrier is now \emph{overdense}---meaning that classical 
transmission is forbidden. For perpendicular incidence, the free `states' are classically 
inaccessible, so their contribution to the eigensum of states forming the total 
wavefunction is exponentially small.  

For states with $\beta \gg 1$, the problem is most easily solved using the WKB 
technique in (actual) configuration space, since there are no turning-points to 
contend with. Constraining the discussion to perpendicular incidence means however 
that only those states with $\beta$ a little greater than one need be calculated, 
so the `close to the separatrix' treatment of the preceeding sections must be 
generalised to encompass $\beta >1$. Since the essentials of the application of 
the uniform method to the Raman-Nath equation have already been conveyed, the 
following treatment is intended to be more of a `recipe' than a detailed account.

The overdense barrier equation will be taken as
\begin{equation}
\frac{d^{2} \phi}{d \sigma^{2}}+ \left(\sigma^{2}-t \right) \phi=0 
\label{eq:overdense-barrier-equation}
\end{equation}
with $t$ a positive quantity. The connection with the parabolic cylinder 
equation (\ref{eq:parabolic-cylinder-equations}) is made with the aid of the 
transformations
\begin{eqnarray}
a  & = & -\mathrm{i} \frac{t}{2}  \\
g & = & \sqrt{2} \sigma \mathrm{e}^{\mathrm{i} \pi /4}.
\end{eqnarray}
To remove any ambiguity regarding the phase momentum function $p_{2}$ for the barrier, 
it will be written as 
\begin{equation}
\bar{p}_{2}= \arccos \left[\beta-y^{2} \right] = 
\left\{ \begin{array}{ll} \mathrm{i} \ \mathrm{arccosh} 
\left[\beta-y^{2} \right] & \mbox{if $0 \leq y \leq \sqrt{\beta-1}$ } \\[2ex]
\pi -\arccos \left[y^{2}-\beta \right] & \mbox{if  $\sqrt{\beta-1} 
\leq y < \sqrt{1+ \beta}$ } \end{array} \right. 
\end{equation}
where the central barrier lies between $\pm \sqrt{\beta-1}$. The actions 
generated from these momenta, using $y=\sqrt{\beta-1}$ as the reference point, are
\begin{equation}
\bar{S}_{0}^{y < \sqrt{\beta-1}} (\sqrt{\beta-1},y,\beta)= \mathrm{i} 
\biggl( y \ \mathrm{arccosh}  \left[\beta-y^2 \right]  + 2 \mathrm{i} 
\sqrt{\beta-1} \mathrm{E} \left. \left( \frac{1}{2} \arccos 
\left[\beta -y^{2}\right] \right| \frac{2}{1-\beta} \right) \biggr)  
\label{eq:overdense-action-inside-barrier}
\end{equation}
and
\begin{equation}
\begin{split}
\bar{S}_{0}^{y > \sqrt{\beta-1}} (\sqrt{\beta-1},y,\beta)= \pi y & + 2\sqrt{\beta+1} 
\mathrm{E} \left. \left( \frac{1}{2} \arccos \left[y^{2}-\beta \right] \right| 
\frac{2}{1+\beta} \right)\\ & - 2 \sqrt{\beta+1}\mathrm{E} \left( \left. \frac{\pi}{2} 
\right| \frac{2}{1+\beta} \right)-
 y \ \arccos  \left[y^2-\beta \right]. 
\end{split}
\end{equation}

As  before, the comparison equation (\ref{eq:overdense-barrier-equation}) gives rise 
to the mapping function by setting 
\begin{equation}
\bar{S}_{0}^{y< \sqrt{\beta-1}}=\int_{\sqrt{t}}^{\sigma} \sqrt{\sigma^{2}-t} \;d 
\sigma =  \mathrm{i} \frac{t}{2} \left( \arcsin \left[\frac{\sigma}{\sqrt{t}} \right] 
+ \frac{\sigma}{\sqrt{t}} \sqrt{1- \frac{\sigma^{2}}{t}} -\frac{\pi}{2} \right) 
\label{eq:overdense-comparison-action-inside-barrier}
\end{equation}
and
\begin{equation}
\bar{S}_{0}^{y>\sqrt{\beta-1}}=\int_{\sqrt{t}}^{\sigma} \sqrt{\sigma^{2}-t} 
\;d \sigma = \frac{t}{2} \left( \frac{\sigma^{2}}{t} 
\sqrt{1-\frac{t}{\sigma^{2}}}-\mathrm{arccosh} 
\left[\frac{\sigma}{\sqrt{t}} \right] \right). 
\label{eq:overdense-comparison-action-outside-barrier}
\end{equation}
In particular, the `barrier integral' which fixes the value of $t$ once $\beta$ 
is known, can this time be conducted along the real axis, and gives, using 
(\ref{eq:overdense-action-inside-barrier}) and 
(\ref{eq:overdense-comparison-action-inside-barrier}),
\begin{equation}
2 \mathrm{i} \sqrt{\Lambda} \sqrt{\beta-1} \mathrm{E} \left. 
\left(\frac{1}{2} \arccos \left[ \beta \right] \right| \frac{2}{1-\beta} 
\right)= -\frac{t \pi}{4}.
\end{equation}

The correct solution to the barrier equation is still the even power series
\begin{equation}
\Theta_{1}(a,g)=\Theta_{1} \left( -\mathrm{i} \frac{t}{2}, \sqrt{2} \sigma 
\mathrm{e}^{\mathrm{i} \pi /4} \right)
\end{equation}
and so the transitional approximation for the overdense barrier becomes
\begin{equation}
\begin{split}
B(y)=(-1)^{n}C(y) & =(-1)^{n} \psi_{\mathrm{transitional}} \\ & \propto 
\quad (-1)^{n} \left( \frac{\sigma^{2}-t}{1-\left(y^{2}-\beta \right)^{2}} 
\right)^{1/4} \mathrm{e}^{-\mathrm{i} \sigma^{2}/2} \; \hspace{1em}_{1} \! F_{1} 
\left(-\mathrm{i}\frac{t}{4}+\frac{1}{4}; \frac{1}{2}; \mathrm{i} \sigma^{2} \right). 
\label{eq:transitional-wavefunction-overdense}
\end{split}
\end{equation}

\textbf{17. Asymptotic matching to the overdense WKB expression}

The transitional wavefunction differs by a few sign changes from the underdense case, 
and for large $\sigma$ these produce the modified oscillatory behaviour:
\begin{equation}
\begin{split}
\Theta_{1} & \biggl( - \mathrm{i}\frac{t}{2}, \sqrt{2} \sigma \mathrm{e}^{\mathrm{i} 
\pi/4} \biggr) \\ & \sim 2 \frac{\Gamma \left( \frac{1}{2} \right)}{\left| \Gamma 
\left(\frac{1}{4}+ \mathrm{i} \frac{t}{4} \right) \right|} \sigma^{-1/2} 
\mathrm{e}^{\pi t/8-t/(4 \sigma^{2})} \cos \left( \frac{t}{2} \ln \sigma - 
\frac{ \sigma^{2}}{2} -\mathrm{Arg} \left[ \Gamma \left(\frac{1}{4} + 
\mathrm{i} \frac{t}{4} \right) \right] +\frac{\pi}{8}+ \mathcal{O} 
\left(\frac{1}{\sigma^{4}} \right) \right). \label{eq:1F1-asymptotic-overdense}
\end{split}
\end{equation}
Expanding the rhs of Equation (\ref{eq:overdense-comparison-action-outside-barrier}) 
for $\sigma \gg t$ gives
\begin{equation}
\frac{t}{2} \left( \frac{\sigma^{2}}{t} \sqrt{1-\frac{t}{\sigma^{2}}}-\mathrm{arccosh} 
\left[\frac{\sigma}{\sqrt{t}} \right] \right) \sim -\frac{t}{2} \ln \sigma + 
\frac{t}{4} \ln t - \frac{t}{2} \ln 2+ \frac{\sigma^{2}}{2} - \frac{t}{4}
\end{equation}
from which one deduces the unknown phase angle $\mu$, appearing in the WKB 
approximation for the overdense barrier (see Equation (\ref{eq:wkb-with-mu})), to be
\begin{equation}
\mu=-\frac{t}{4} \ln t + \frac{t}{2} \ln 2 + \frac{t}{4} + \mathrm{Arg} \left[\Gamma 
\left(\frac{1}{4}+ \mathrm{i} \frac{t}{4} \right) \right] - \frac{\pi}{8}.
\end{equation}

\textbf{18. The overdense eigenvalues}

Once again $\mu$ can be successfully employed in the accurate determination of the 
eigenvalues $\beta$. Following the empirical observations from the underdense case, 
the WKB expression emanating from the outer turning-point and that from the inner 
turning-point are matched at a point $y$ corresponding to one of the beams. This 
time the choice of $y=0$ is not available since only the phase for the WKB 
approximation outside the barrier is known. The next most obvious choice is either 
the inner or outer turning-point since there the phase of the WKB expressions are 
simplest, but in general these classically determined points will not fall on a 
diffracted beam. Selecting a random beam, $y=m/\sqrt{\Lambda}$, with $m$ an integer, 
giving of value $y$ lying between the two turning-points, will suffice. The condition 
giving the permitted values of $\beta$ for even eigenstates then alternates between
\begin{equation}
\begin{split}
\cos \biggl( \sqrt{\Lambda}\bar{S}_{0}^{y>\sqrt{\beta-1}} \left(\sqrt{\beta-1},
\frac{m}{\sqrt{\Lambda}},\beta \right) & + \mu(\beta)+\sqrt{\Lambda} \pi 
\frac{m}{\sqrt{\Lambda}} \biggr) \\ & - \cos \left( \sqrt{\Lambda} S_{0} 
\left(\sqrt{1+\beta}, \frac{m}{\sqrt{\Lambda}},\beta \right) + \frac{\pi}{4} \right)=0
\end{split}
\end{equation} 
and
\begin{equation}
\begin{split}
\cos \biggl( \sqrt{\Lambda}\bar{S}_{0}^{y>\sqrt{\beta-1}} \left(\sqrt{\beta-1},
\frac{m}{\sqrt{\Lambda}},\beta \right) & + \mu(\beta)+\sqrt{\Lambda} \pi 
\frac{m}{\sqrt{\Lambda}} \biggr) \\ & - \cos \left( \sqrt{\Lambda} S_{0} 
\left(\sqrt{1+\beta}, \frac{m}{\sqrt{\Lambda}},\beta \right) + \frac{5 \pi}{4} 
\right)=0.
\end{split}
\end{equation}
Both of these equations have zeros which do not correspond to the eigenvalues, 
the correct ones being those for which gradient of the lhs' are positive 
(this is the opposite of the underdense case).
As before, the accuracy which is achieved gives confidence to the method: for 
$\Lambda=12500$ the first two free eigenvalues given by numerical diagonalisation 
are $\beta=1.003356$ and $\beta=1.012155$, for which this WKB matching technique 
gives $\beta=1.003358$ and $\beta=1.012156$ respectively. 

\textbf{19. The overdense eigenvectors}

\begin{figure}[htbp]
\begin{center}
\centerline{\epsfig{figure=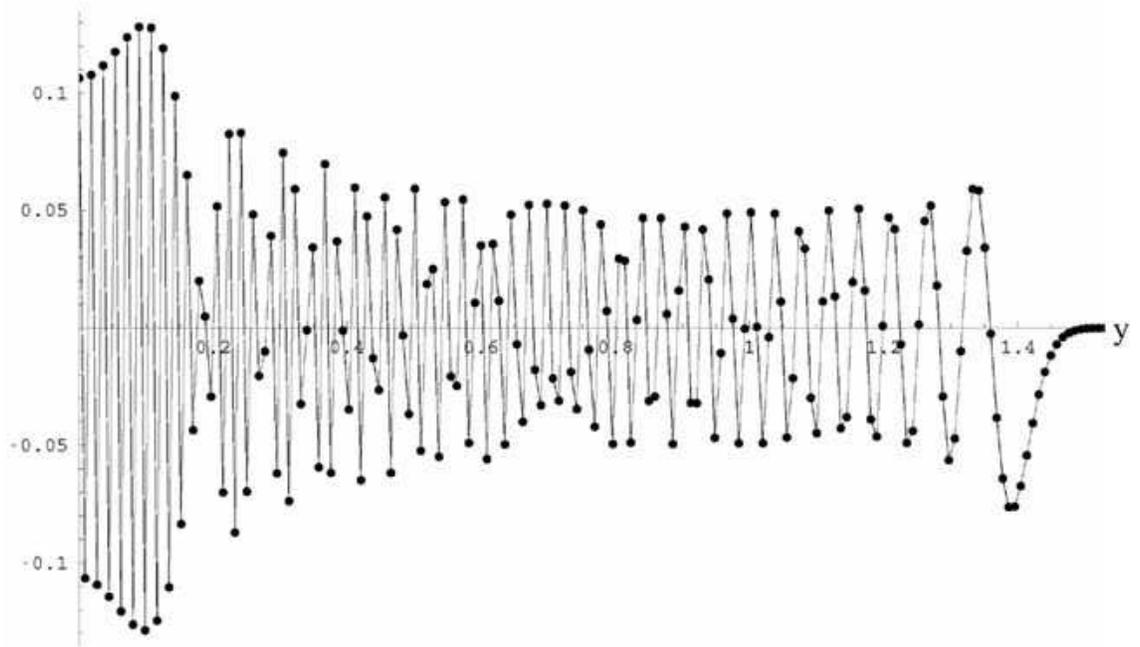,height=15cm,angle=-90}}
\end{center}
\caption{The $201^{\mathrm{st}}$ Bloch wave, which is 
the first `free' eigenvector. This requires both the overdense parabolic 
barrier solution, and an Airy function as transitional approximations. 
The two 
are joined at the $83^{\mathrm{rd}}$ diffracted beam, which is at $y=0.742$. 
The dots are the purely numerical calculation.} 
\label{fig:freeevec}
\end{figure}

Knowing the value of $\beta$, one is in a position to calculate the transitional 
approximation to the overdense eigenvector
\begin{equation}
B_{\mathrm{barrier}}(y)=(-1)^{n} \mathcal{N} \frac{\left| \Gamma \left(\frac{1}{4}+ 
\mathrm{i} \frac{t}{4} \right) \right| \mathrm{e}^{-\pi t/8}}{2 \Gamma \left( 
\frac{1}{2} \right)} \left( \frac{\sigma^{2}-t}{1-\left(y^{2}-\beta \right)^{2}} 
\right)^{1/4} \mathrm{e}^{-\mathrm{i} \sigma^{2}/2} \; \hspace{1em}_{1} \! F_{1} 
\left(-\mathrm{i}\frac{t}{4}+\frac{1}{4}; \frac{1}{2}; \mathrm{i} \sigma^{2}\right). 
\label{eq:overdense-wavefunction}
\end{equation}
The Airy function approximation for the outer turning-point remains the same as before. 
Figure \ref{fig:freeevec} shows the first free eigenvector made up of the overdense 
barrier and Airy function approximations.

\textbf{20. Conclusion}

The diffraction of a plane wave by a sinusoidal potential is conveniently
described by the Raman-Nath equation (Mathieu equation in Fourier space).
A  method for calculating the eigenvalues (characteristic values) and 
continuized eigenvectors  of this differential difference equation 
 in the short 
wavelength limit is given. Working in Fourier space circumvents some of
the difficulties associated with the infinite number of 
turning points inherent in a periodic potential. 

WKB-type solutions to the Raman-Nath equation serve as a starting point. 
Eigenvalues then follow from a simple Bohr-Sommerfeld relation. 
Whilst the WKB solutions to the Raman-Nath equation still contain divergences, 
they reveal that the Raman-Nath equation can be interpreted as describing a wave 
in a double well potential, for which simple uniform approximations---solutions 
without singularities---exist in terms of the parabolic cylinder functions.
There are three situations. Firstly, the `bound' eigenstates lying below the 
separatrix in coordinate space lie above the central barrier in the double well 
in Fourier space and so a single uniform approximation in terms of Hermite 
polynomials suffices. Secondly, at or just below the coordinate space separatrix 
the eigenstates lie at or just above the central barrier in the double well in 
Fourier space, causing two new turning points to appear. A complete eigenfunction 
with no singularities can be constructed by smoothly sewing together two transitional 
approximations---a parabolic cylinder function and an Airy function (the double well 
in Fourier space is symmetrical so only two transitional approximations are required 
rather than three). Thirdly, above the coordinate space separatrix and consequently
below the central barrier in Fourier space, the `free' eigenstates are also given by 
matching a parabolic cylinder function and an Airy function. A perscription is given for 
the modification of the Bohr-Sommerfeld rule for the eigenvalues near the separatrix.

When considering the diffraction of waves by a sinusoidal 
potential, knowing the eigenfunctions of the potential allows one to propagate the 
incident wave for any interaction distance and hence investigate dynamical diffraction 
phenomena which go beyond the phase-grating/Raman-Nath approximation, such as caustics 
(natural focussing). Semiclassically, the superposition of eigenfunctions giving the 
diffraction pattern due to an incident plane wave only contains an exponentially small 
contribution from the free eigenstates. However, as $\hbar \rightarrow 0$ ($\Lambda 
\rightarrow \infty$), even the number of `bound' states become infinite. In a companion 
paper (O'Dell 2001) it will be demonstrated that a Poisson 
resummation of the eigenfunction superposition 
produces a new series each term of which is associated with \emph{classical paths} belonging 
to a different topological class.  Furthermore, the number of terms required in this new sum 
depends linearly on the distance propagated through the potential (independent of the size 
of $\hbar$)---only a finite number of terms are required for finite propagation distances, 
and so is computationally superior to the original eigenfunction sum which requires an 
infinite number of terms in the $\hbar \rightarrow 0$ limit.

\textbf{21. Acknowledgements}

It is a pleasure to thank M.V. Berry, J.H. Hannay, W. Schleich and V.P. Yakovlev for 
numerous discussions and suggestions. I also thank the University of Bristol, U.K., 
for a studentship during which this work was undertaken.

\newpage
\textbf{Appendix: the method of uniform approximation}

Details can be found in the review by Berry and Mount (1972). 
The objective is to obtain an approximate solution of the Helmholtz equation
\begin{equation}
\frac{d^{2} \psi(q)}{d q^{2}} + \chi(q) \psi(q)=0 \label{eq:psi-eqn}
\end{equation}
in terms of solutions to one of the `studied' equations, which will be written
\begin{equation}
\frac{d^{2} \phi(\sigma)}{d \sigma^{2}} + \Gamma(\sigma) \phi(\sigma)=0. 
\label{eq:phi-eqn}
\end{equation} 
The choice of studied equation is determined by $\Gamma(\sigma)$ (not \emph{the} 
Gamma function) being in some way similar to $\chi(q)$. This similarity implies 
that $\phi(\sigma)$ also resembles the wavefunction $\psi(q)$, and ``can be changed 
into it by stretching or contracting it a little and changing the amplitude a 
little''. And so $\psi(q)$ will be expressed in terms of $\phi(\sigma)$ 
\begin{equation}
\psi(q)=f(q) \phi(\sigma(q)).
\end{equation}
Substitution of this definition into (\ref{eq:psi-eqn}) and making use 
of (\ref{eq:phi-eqn}) leaves
\begin{equation}
\frac{d^{2}f}{d q^{2}}+ \chi f \phi-f \left(\frac{d \sigma}{d q} \right)^{2} 
\Gamma \phi+ \frac{d \phi}{d \sigma} \left(2 \frac{df}{dq}\frac{d \sigma}{dq}+
f\frac{d^{2} \sigma}{d q^{2}} \right) = 0. 
\label{eq:uniform-generator}
\end{equation}
The amplitude $f(q)$ is as yet unspecified, so it is chosen 
to simplify (\ref{eq:uniform-generator}) as much as possible. Putting
\begin{equation}
f=\left(\frac{d \sigma}{d q} \right)^{-\frac{1}{2}} \label{eq:f-definition}
\end{equation}
renders (\ref{eq:uniform-generator}) into an equation purely for the  `mapping 
function' $\sigma(q)$
\begin{equation}
\chi=\left(\frac{d \sigma}{d q} \right)^{2} \Gamma - \left(\frac{d \sigma}{dq} 
\right)^{\frac{1}{2}} \frac{d^{2}}{d q^{2}} \left( \frac{d \sigma}{dq} 
\right)^{-\frac{1}{2}}
\end{equation}
which, when solved, gives $\sigma$ as a function of $q$.
If a good choice of comparison function $\Gamma(\sigma)$ has been made, then 
$\sigma(q)$ will be a slowly varying function and the second term on the rhs 
of (\ref{eq:uniform-generator}) will be much smaller than the first. Clearly 
the criterion for this to be the case is
\begin{equation}
\epsilon (q) \equiv \left| \frac{1}{\chi(q)} \left( \frac{d \sigma}{dq} 
\right)^{\frac{1}{2}} \frac{d^{2}}{d q^{2}} \left( \frac{d \sigma}{dq} 
\right)^{-\frac{1}{2}} \right| \ll 1.
\end{equation}
When this is satisfied, the mapping relation reduces to
\begin{equation}
\frac{d \sigma}{d q} \simeq \left( \frac{\chi(q)}{\Gamma (\sigma)} 
\right)^{\frac{1}{2}} \label{eq:dsigma/dq}
\end{equation}
which through definition (\ref{eq:f-definition}) also gives the amplitude 
$f$.
Thus, by picking two points $\sigma_{0}$ and $q_{0}$ which are `equivalent',  
one finds $\sigma(q)$ from
\begin{equation}
\int_{\sigma_{0}}^{\sigma}\sqrt{\pm \Gamma(\sigma)} \; d \sigma = 
\int_{q_{0}}^{q}\sqrt{\pm \chi(q)} \; d q \label{eq:int[gamma]dsigma=int[chi]dq}
\end{equation}
where the $+$ or the $-$ version can be chosen depending on the situation. 
The approximate solution to (\ref{eq:psi-eqn}) is then
\begin{equation}
\psi(q) \simeq \left( \frac{\Gamma \left(\sigma (q) \right)}{\chi (q)} 
\right)^{\frac{1}{4}} \phi \left( \sigma (q) \right).
\end{equation}

In order for the comparison method to be viable, the mapping from $q$ to 
$\sigma$ must be one to one, which requires that $d \sigma/dq$ is never zero 
or infinite. Examining (\ref{eq:dsigma/dq}) this means that $\chi$ and $\Gamma$ 
must not diverge---which is assumed to be the case---and more relevantly, their 
zeros must be made to correspond. The zeros are of course the turning-points, 
and so, as Berry and Mount emphasise, ``in the semiclassical limit all problems 
are equivalent which have the same \emph{classical turning-point structure}''.  

Perhaps the best known example of the uniform approximation is for the lone, 
first order (that is, the potential is locally  linear) turning-point leading 
to the comparison equation
\begin{equation}
\frac{d^{2} \sigma}{d \sigma^{2}}- \sigma \phi = 0 \label{eq:airy-comparison-equation}
\end{equation}
whose solution is the Airy function, Ai($\sigma$). Many potentials of interest 
are linear close to the turning point. As one moves away from the turning point 
the Airy function can be smoothly matched onto a WKB solution which is capable 
of handling very complicated potentials provided there are no turning points. 
Used in this way, as a patch across the turning point, the Airy function 
constitutes what is sometimes referred to as a \emph{transitional} approximation. 


\newpage

\begin{thebibliography}{30}

\bibitem{a+s}
Abramowitz M and Stegun I A 1964 \textit{Handbook of Mathematical Functions}
(Washington, DC: National Bureau of Standards)

\bibitem{adams}
Adams C S, Sigel M and Mlynek J 1994 Atom optics \textit{Phys. Rep.}
\textbf{240} 143--210

\bibitem{ancey}
Ancey S, Folacci A and Gabrielli P  2000
Exponentially improved asymptotic expansions for resonances
of an elliptic cylinder 
\textit{J. Phys. A: Math. Gen.} \textbf{33} 3179--3208

\bibitem{mvb:thesis}
Berry M V 1966 \textit{The Diffraction of Light by Ultrasound}
(New York: Academic)

\bibitem{mvb:high-energy}
Berry M V 1971
Diffraction in crystals at high energies
\textit{J. Phys. C: Solid State Phys.} \textbf{4} 697--722

\bibitem{berry+mount}
Berry M V and Mount K E 1972
Semiclassical approximations in wave mechanics
\textit{Rep. Prog. Phys.} \textbf{35} 315--397

\bibitem{mvb:leshouches80}
Berry M V 1981 \textit{Les Houches Lecture Series} session XXXV,
ed R. Balian \textit{et al.} (Amsterdam: North Holland) pp 455--541


\bibitem{berry+odell}
Berry M V and O'Dell D H J 1998
Diffraction by volume gratings with imaginary potentials
\textit{J. Phys. A: Math. Gen.} \textbf{31} 2093--2101

\bibitem{berry+odell99}
Berry M V and O'Dell D H J 1999 Ergodicity in wave-wave diffraction
\textit{J. Phys. A: Math. Gen.} \textbf{32} 3571--3582

\bibitem{brillouin:21}
Brillouin L 1921 \textit{Annal Physique} \textbf{17} 103


\bibitem{book:atom-photon}
Cohen-Tannoudji C, Dupont-Roc J and Grynberg G 1992
\textit{Atom-Photon Interactions} (New York: Wiley-Interscience)

\bibitem{dingle67i}
Dingle R B and Morgan G J 1967a
{WKB} methods for difference equations {I}
\textit{Applied Scientific Research} \textbf{18} 221--237

\bibitem{dingle67ii}
Dingle R B and Morgan G J 1967b
{WKB} methods for difference equations {II}
\textit{Applied Scientific Research} \textbf{18} 238--245

\bibitem{gould:expt86}
Gould P L, Ruff G A and Pritchard D E 1986
Diffraction of Atoms by Light: The Near-Resonant
{K}apitza-{D}irac Effect \textit{Phys. Rev. Lett.}
\textbf{56} 827--830

\bibitem{g+r}
Gradshteyn I S and  Ryzhik I M 1965
\textit{Table of Integrals, Series and Products}
(New York: Academic)


\bibitem{book:mechanical-action-light}
Kazantsev A P, Surdutovich G I and Yakovlev V P 1991
\textit{Mechanical Action of Light on Atoms}
(Singapore: World Scientific)

\bibitem{odell:thesis}
O'Dell D H J 1999 \textit{The Diffraction of Atoms by Light}
(University of Bristol PhD thesis: unpublished)

\bibitem{odell:tobepublished}
O'Dell D H J 2001 to be published

\bibitem{olver}
Olver F W J 1954
The asymptotic solutions of linear differential equations of
the second order for large values of a parameter
\textit{Phil. Trans. R. Soc.} A \textbf{247} 307-27


\bibitem{raman-nath:35i}
Raman C V and Nagendra Nath N S 1935
The Diffraction of Light by High Frequency Sound Waves: Part {I}
\textit{Proc. Indian Acad. of Sci.} A \textbf{2} 406--412

\bibitem{raman-nath:36ii}
Raman C V and Nagendra Nath N S 1936
The Diffraction of Light by High Frequency Sound Waves: Part
{IV} \textit{Proc. Indian Acad. of Sci.} A \textbf{3} 119--25


\bibitem{rasel:expt95}
Rasel E M, Oberthaler M K, Batelaan H, Schmiedmayer J and
Zeilinger A 1995
Atom Wave Interferometry with Diffraction Gratings of Light
\textit{Phys. Rev. Lett.} \textbf{75} 2633--2637
 
\bibitem{sanders}
Sanders F H 1936 \textit{Canadian J. Research} \textbf{A14}
158

\bibitem{sieber}
Sieber M 1997
Semiclassical transition from an elliptic to an oval billiard 
\textit{J. Phys. A: Math. Gen.} \textbf{30} 4563--4596

\bibitem{waalkens}
Waalkens H, Wiersig J and Dullin H R 1997
Elliptic quantum billiard
\textit{Ann. Phys.} \textbf{260} 50--90


\bibitem{yakovlev:private}
Yakovlev V P 1997 private communication


\end{thebibliography}

\newpage

\begin{table}[htbp]
\begin{center}
\begin{tabular}{||c|c|c|c|c||} \hline
$\Lambda$ & $j$ & fully numerical & single well calc. & modified calc. \\ \hline
 12500 & 200 & 0.996129 & 0.996824 & 0.996131 \\
 & 198 & 0.987197 & 0.987337 & 0.987199 \\
 & 196 & 0.976711 & 0.976759 & 0.976713 \\
 & 194 & 0.965430 & 0.965461 & 0.965432 \\
 & 192 & 0.953575 & 0.953599 & 0.953578 \\
 & 190 & 0.941244 & 0.941264 & 0.941247 \\
 & 188 & 0.928498 & 0.928515 & 0.928499 \\
 & 186 & 0.915379 & 0.915394 & 0.915381 \\ \hline
250000 & 900 & 0.999954 & -- & 0.999954  \\ \hline
\end{tabular}
\end{center}
\caption{The bound eigenvalues near the separatrix: 
comparison of numerical result with the standard Bohr-Sommerfeld condition for a well 
(\ref{eq:bohr-sommerfeld}), and the modified conditions 
(\ref{eq:underdense-eigenvalue-condition-up})--(\ref{eq:underdense-eigenvalue-condition-down}).}
\label{tab:evalues}
\end{table}

\newpage

)

\end{document}